\documentclass[twocolumn,aps,showpacs,prb,tightenlines,amsmath,amssymb]{revtex4}
\usepackage{bm}
\usepackage{graphicx}              
\usepackage{amssymb}               
\usepackage{amsmath}                     
\usepackage{colordvi}
\usepackage{calrsfs} 
\usepackage[thinlines,thiklines]{easybmat}
\newcommand{\bgreek}[1]{\mbox{\boldmath$#1$\unboldmath}}
\begin{document}   

\title{Spin diffusion in ultracold spin-orbit coupled $^{40}$K gas}
 
\author{T. Yu}
\author{M. W. Wu}
\thanks{Author to whom correspondence should be addressed}
\email{mwwu@ustc.edu.cn.}
\affiliation{Hefei National Laboratory for Physical Sciences at
  Microscale,  Key Laboratory of Strongly-Coupled Quantum Matter Physics and Department of Physics, 
University of Science and Technology of China, Hefei,
  Anhui, 230026, China} 
\date{\today}

\begin{abstract} 
We investigate the steady-state spin diffusion for ultracold spin-orbit coupled
  $^{40}$K gas by the kinetic spin
Bloch equation approach both analytically and
numerically. Four configurations, i.e., the spin diffusions along two specific directions 
with the spin polarization perpendicular (transverse configuration) and
parallel (longitudinal configuration) to the effective 
Zeeman field are studied. It is found that the behaviors of the steady-state spin diffusion  
 for the four configurations are very different, which
are determined by three characteristic lengths: the mean free path $l_{\tau}$, the Zeeman oscillation
length $l_{\Omega}$ and the spin-orbit coupling oscillation length
$l_{\alpha}$. It is analytically revealed and numerically confirmed
that by tuning the scattering strength, the system can be divided into {\it five} regimes: I, weak
scattering regime ($l_{\tau}\gtrsim l_{\Omega}, 
l_{\alpha}$); II, Zeeman field-dominated moderate scattering regime ($l_{\Omega}\ll
l_{\tau}\ll l_{\alpha}$); III, spin-orbit coupling-dominated moderate scattering regime ($l_{\alpha}\ll
l_{\tau}\ll l_{\Omega}$); IV, relatively strong scattering regime ($l_{\tau}^c\ll l_{\tau}\ll l_{\Omega},
l_{\alpha}$); V, strong scattering regime ($l_{\tau}\ll l_{\Omega},
l_{\alpha},l_{\tau}^c$), with $l_{\tau}^c$ representing the crossover length
 between the relatively strong and strong scattering
regimes.
 In different regimes, the behaviors of
 the spacial evolution of the steady-state spin polarization are very 
rich, showing different dependencies on the scattering strength,
Zeeman field and
spin-orbit coupling strength. The rich behaviors of the spin diffusions in different regimes
 are hard to be understood in the framework of the simple drift-diffusion
  model or the direct inhomogeneous
  broadening picture in the literature. However, almost all these rich behaviors can be
  well understood from our {\it modified}
drift-diffusion model and/or {\it modified} inhomogeneous broadening
picture. Specifically, several anomalous features of the spin
diffusion are revealed, which are in contrast to those obtained from {\it both} the simple drift-diffusion
  model and the direct inhomogeneous
  broadening picture.         
\end{abstract}

\maketitle
\section{Introduction}
In recent years, spin dynamics including spin relaxation and spin
diffusion/transport is extensively studied in both 
Bose\cite{Bose_oscillation_1,Bose_oscillation_2,Bose_3,Bose_4,Bose_5,Bose_spin_1,Bose_SOC,Spielman_2015} and
 Fermi\cite{Fermi_tunneling_1,Fermi_tunneling_2,Fermi_tunneling_3,Universe_1,Universe_2,
Universe_3,Universe_4,Universe_5,Leggett,Universe_Leggett_1,Universe_Leggett_2,Universe_Leggett_3,
Universe_Leggett_4,Spin_segregation_1,Spin_segregation_2,Spin_segregation_3,
Spin_segregation_4,Fermi_SOC_1,Fermi_SOC_2,
relaxation_1,relaxation_2,relaxation_3,relaxation_4,relaxation_5}
 cold atoms. For the Bose system, the spin dynamics of the Bose-Einstein condensation
  has attracted much attention.\cite{Bose_oscillation_1,
Bose_oscillation_2,Bose_3,Bose_4,Bose_5,Bose_spin_1,Bose_SOC,Spielman_2015} For the Fermi
cold atoms, the systems
 without\cite{Fermi_tunneling_1,Fermi_tunneling_2,Fermi_tunneling_3,Universe_1,Universe_2,
Universe_3,Universe_4,Universe_5,Leggett,Universe_Leggett_1,Universe_Leggett_2,Universe_Leggett_3,
Universe_Leggett_4,Spin_segregation_1,Spin_segregation_2,Spin_segregation_3,
Spin_segregation_4} and with\cite{Fermi_SOC_1,Fermi_SOC_2,
relaxation_1,relaxation_2,relaxation_3,relaxation_4,relaxation_5}
spin-orbit coupling (SOC) are extensively investigated. In the
  absence of the SOC, many interesting phenomena, such as the Leggett-Rice effect in
unitary
 gas\cite{Leggett,Universe_Leggett_1,Universe_Leggett_2,Universe_Leggett_3,
Universe_Leggett_4} and anomalous
  spin segregation in extremely weak scattering limit,\cite{Spin_segregation_1,Spin_segregation_2,Spin_segregation_3,
Spin_segregation_4} have enriched the understanding of the spin dynamics of 
Fermions. With the synthetic SOC experimentally realized by laser
 control technique in cold atoms,\cite{Bose_SOC,Fermi_SOC_1,Fermi_SOC_2,Spielman_2015}
 the spin relaxation for the Fermi cold atoms with SOC has been studied
  both experimentally\cite{Fermi_SOC_1,Fermi_SOC_2,relaxation_5} and
 theoretically.\cite{relaxation_1,relaxation_2,relaxation_3,relaxation_4} 
This is partly motivated by the well-controlled
 laser technique, which provides more freedom for the cold atoms than the
 conventional solids.
 On one hand, rich regimes can be realized by tuning the SOC strength; on the
 other hand, not only the interatom interaction can be tuned by
the Feshbach resonance,\cite{Feshbach} but also the atom-disorder interaction
can be introduced and controlled by the speckle laser
 technique.\cite{Disorder_3D_Fermion,Disorder_3D_Boson,Disorder_1D_Boson,Disorder_theory}

The experimentally realized effective Zeeman field and SOC
 provide an effective magnetic field,
 which reads\cite{Bose_SOC,Fermi_SOC_1,Fermi_SOC_2} 
\begin{equation}
{\bf{\Omega}}({\bf{k}})=(\Omega,0,\delta+\alpha k_x).
\label{SOC}
\end{equation}
In above equation, ${\bf k}=(k_x,k_y,k_z)$ denotes the center-of-mass momentum of the atom;
 $\Omega$ acts as an effective Zeeman field along the $\hat{\bf x}$-direction; $\delta$ is the Raman detuning,
 which is set to be zero in our work;  
$\Omega_z({\bf k})=\alpha k_x$ represents the ${\bf k}$-dependent effective
magnetic field along the 
$\hat{\bf z}$-direction, which is perpendicular to the Zeeman field, with $|\alpha|$ being
 the strength of the spin-orbit coupled field.  With this specific  
 effective magnetic field ${\bf{\Omega}}({\bf{k}})$ by setting
 $\delta=0$, it has been revealed that both the
 conventional\cite{relaxation_1,relaxation_2,relaxation_4} and
anomalous\cite{Anomalous_DP_1,relaxation_3,Anomalous_DP_2} D'yakonov-Perel' (DP)\cite{DP} spin relaxations can be
realized with $\langle|\Omega_z({\bf k})|\rangle\gtrsim \Omega$ and
$\langle|\Omega_z({\bf k})|\rangle\ll \Omega$, respectively. For the
conventional situation,
 in the strong (weak) scattering limit when $\langle|\Omega({\bf
  k})|\rangle\tau_k^\ast\ll 1$ ($\langle
|\Omega({\bf k})|\rangle\tau_k^\ast\gtrsim 1$), 
the spin relaxation time (SRT) $\tau_s$ is inversely
proportional (proportional) to the momentum scattering time
$\tau_k^{\ast}$. $\langle...\rangle$ here denotes the ensemble average. 
For the anomalous situation,\cite{Anomalous_DP_1,relaxation_3,Anomalous_DP_2} it
has been found that by tuning the
   interatom interaction, the transverse spin
relaxation can be divided into four regimes: the normal weak
scattering ($\tau_{\rm s}\propto \tau_k^{\ast}$),
 the anomalous DP-like ($\tau_{\rm s}^{-1}\propto \tau_k^{\ast}$),
the anomalous Elliott-Yafet (EY)-like\cite{Yafet,Elliot} ($\tau_{\rm s}\propto \tau_k^{\ast}$)
and the normal strong scattering ($\tau_{\rm s}^{-1}\propto
\tau_k^{\ast}$) regimes. Whereas the longitudinal spin relaxation can be divided
into two: i.e., the anomalous EY-like and the normal strong scattering
regimes.\cite{Anomalous_DP_1}

In contrast to the spin relaxation, the study for the spin
diffusion in cold atoms with SOC has not yet been reported in the literature.
 However, the experimental configuration 
  realized by Brantut {\em et al.},\cite{Configuration}
 for the ``charge'' diffusion of cold atoms can be applied to the spin diffusion
  straightforwardly. In their experiment,
 by adding the ``barrier'' laser in the
    middle of the cold atoms, the system is separated to the left and right parts.
 In the left part,  
the spin-polarized cold atoms
  can be prepared;\cite{Fermi_SOC_1,Fermi_SOC_2,relaxation_4,relaxation_5} 
 whereas in the right part,
 the system remains in the equilibrium
  state. Specifically, the atom densities in the 
  left and right parts are prepared to be the same. With the SOC introduced to
  the right part,\cite{Fermi_SOC_1,Fermi_SOC_2,relaxation_4,relaxation_5} i.e., the spin diffusion
 region, by removing the ``barrier'' laser, 
  this configuration can be used to study spin
diffusion along one direction for the three dimensional (3D) Fermi gas with
SOC.

 In the coordinate defined in Eq.~(\ref{SOC}), there are two specific 
  configurations with the spin diffusions along the $\hat{\bf x}$- and
  $\hat{\bf y}$-directions, respectively. Accordingly, in the scattering-free 
  situation,  
  the ${\bf k}$-dependent spin precession
  frequencies in the spacial domain, i.e., the inhomogeneous
  broadening,\cite{2001,wu-review} in the steady-state spin diffusion along the
  $\hat{\bf x}$-
  and $\hat{\bf y}$-directions are determined by
\begin{eqnarray}
&&{\bgreek{\omega}}^x({\bf{k}})=m{\bgreek \Omega}({\bf k})/k_x=\Big(m\Omega/k_x,0,m\alpha\Big),
\label{diffusion_kx}\\
&&{\bgreek{\omega}}^y({\bf{k}})=m{\bgreek \Omega}({\bf k})/k_y=\Big(m\Omega/k_y,0,m\alpha k_x/k_y\Big),
\label{diffusion_ky}
\end{eqnarray}
respectively.\cite{double,diffusion_k}
Here, $m$ is the atom mass. From
Eq.~(\ref{diffusion_kx}) [Eq.~(\ref{diffusion_ky})], it can be seen that in
contrast to the spin relaxation in the time domain, in inhomogeneous
broadening in the spacial domain, the original ${\bf k}$-independent
Zeeman field $\Omega$ becomes ${\bf k}$-dependent, whereas the ${\bf k}$-dependent
spin-orbit coupled field $m\alpha k_x$ becomes ${\bf
  k}$-independent (remains ${\bf
  k}$-dependent). Hence, for the spin diffusion along the $\hat{\bf x}$-direction, the
inhomogeneous broadening ${\bgreek{\omega}}^x({\bf{k}})$ is similar to the one for spin relaxation in time
domain ${\bgreek \Omega}({\bf k})$ with one
${\bf k}$-independent magnetic field perpendicular to another ${\bf k}$-dependent
one. Accordingly, rich regimes may exit in the steady-state spin diffusion along
the $\hat{\bf x}$-direction for both the spin polarizations perpendicular and parallel to
the Zeeman field. Whereas for the spin diffusion along the $\hat{\bf y}$-direction, from
the different inhomogeneous broadenings in Eqs.~(\ref{diffusion_kx}) and
(\ref{diffusion_ky}), 
its behavior should be very different from the one along the $\hat{\bf x}$-direction.

In the present work, we investigate the steady-state spin diffusion for the
  3D ultracold spin-orbit coupled
  $^{40}$K gas by the kinetic spin
Bloch equation (KSBE) approach\cite{wu-review} both analytically and
numerically. Four configurations, i.e., the spin diffusion along the $\hat{\bf x}$- and
$\hat{\bf y}$-directions for the spin polarization ${\bf P}$ perpendicular
(${\bf P}\parallel \hat{\bf z}$, transverse configuration) and
parallel (${\bf P}\parallel \hat{\bf x}$, longitudinal
  configuration) to the
Zeeman field are studied. It is 
shown analytically that the behaviors of the steady-state spin diffusion 
 for the four configurations are very different, which
are determined by three characteristic lengths:
 the mean free path $l_{\tau}=k\tau_k/m$, the Zeeman oscillation
length $l_{\Omega}={k}/({\sqrt{3}m\Omega})$ and the SOC oscillation
 length $l_{\alpha}={1}/({m|\alpha}|)$. The spin diffusion lengths for the
   spin diffusions in the four configurations are derived in the strong scattering regime,
   which are then extended to the weak scattering one.
 We further find that by dividing the system into different regimes,
 the complex analytical results can be reduced to extremely simple forms.
 It is revealed 
that by tuning the scattering, the system can be divided into {\it five} regimes: I, weak
scattering regime ($l_{\tau}\gtrsim l_{\Omega}, 
l_{\alpha}$); II, Zeeman field-dominated moderate scattering regime ($l_{\Omega}\ll
l_{\tau}\ll l_{\alpha}$); III, SOC-dominated moderate scattering regime ($l_{\alpha}\ll
l_{\tau}\ll l_{\Omega}$); IV, relatively strong scattering regime ($l_{\tau}^c\ll l_{\tau}\ll l_{\Omega},
l_{\alpha}$); V, strong scattering regime ($l_{\tau}\ll l_{\Omega},
l_{\alpha}, l_{\tau}^c$). Here, $l_{\tau}^c$ represents the crossover length 
between the relatively strong and strong scattering regimes.  
In different regimes, the behaviors of the spacial evolution of
the
 steady-state spin polarization are very rich, showing different dependencies on the
scattering strength, Zeeman field and SOC strength. These dependencies are summarized in
Tables~{\ref{table_x}} and {\ref{table_y}} for the spin diffusions along the
$\hat{\bf x}$- and $\hat{\bf y}$-directions, respectively.

 The rich behaviors of the spin diffusions in different regimes
 are hard to be understood in the framework of the previous simple drift-diffusion
  model\cite{Peng43,Peng44,Peng45,Peng46,Peng47,Peng} or the direct inhomogeneous
  broadening [Eqs.~(\ref{diffusion_kx}) and (\ref{diffusion_ky})]
 picture\cite{2001,wu-review,diffusion_k,double,JinLuo_infinite} in the literature. In the simple drift-diffusion model, there are only
  two rather than {\it five} 
  regimes: the strong
 scattering regime with $l_s\propto 1/(m|\alpha|)$ and the weak scattering
 regime with $l_s\propto k\sqrt{\tau_k}/m$. In the present work,
 it is found that the behaviors of the spin diffusions can be analyzed in the
  situation either with strong Zeeman and weak spin-orbit coupled fields
  (Regimes II and V) or weak Zeeman and strong spin-orbit coupled fields
  (Regimes III and IV). Accordingly, our previous inhomogeneous broadenings
  [Eqs.~(\ref{diffusion_kx}) and (\ref{diffusion_ky})] should be extended to the {\it
    effective} ones.
 It is found that when the spin
polarization is parallel to the larger field between the Zeeman and
spin-orbit coupled fields, the spin polarization cannot precess around the effective
inhomogeneous broadening fields efficiently. In this situation, the previous drift-diffusion model is
applicable but $\tau_s({\bf k})$ modified as follows ({\it modified} drift-diffusion
model). In the strong scattering regime, $\tau_s({\bf k})$ remains the SRT in the
   conventional DP mechanism;\cite{relaxation_1,relaxation_2,relaxation_4,wu-review}
 whereas in the moderate scattering regime, $\tau_s({\bf k})$ is replaced by
 the helix spin-flip rates determined in the helix
  space.\cite{relaxation_3,Anomalous_DP_2} When the spin
polarization is perpendicular to the larger field between the Zeeman and
spin-orbit coupled fields, the spin polarization can rotate around the effective
inhomogeneous broadening fields efficiently. Hence, the behavior of the spin diffusion is
determined by the {\it effective}
inhomogeneous broadenings together with the spin-conserving scattering
 ({\it modified} inhomogeneous broadening picture). 
 Based on the modified drift-diffusion model
 and modified inhomogeneous broadening picture, apart from Regime
IV, all the features in
different regimes can be well obtained.

Several anomalous features of the spin
diffusion, which are in contrast to those obtained from {\it both} the simple drift-diffusion
  model and the direct inhomogeneous
  broadening picture, are revealed. In the scattering strength dependence,
 it is found that when $l_{\alpha}\ll l_{\Omega}$,
 the longitudinal spin diffusion along the $\hat{\bf y}$-direction is {\it robust} against the scattering 
in a wide range including both the strong and {\it weak} scattering regimes. 
In the Zeeman field
   dependence, when the system is in
Regime II, the {\it longitudinal} spin diffusion is  
enhanced by the Zeeman
field. In the SOC strength dependence, we find that the spin diffusion length
can be also enhanced by the SOC in Regime III. All
these anomalous behaviors have been well understood from our modified
drift-diffusion model and/or modified inhomogeneous broadening picture.

This paper is organized as follows. In Sec.~{\ref{Model}}, we set
up the model and KSBEs. In
Sec.~{\ref{analytical}},
 we show the analytical results for the transverse and longitudinal spin
   diffusion along the $\hat{\bf x}$- and $\hat{\bf y}$-directions.
 Different dependencies of the spin diffusion length on the mean free path,
 Zeeman oscillation
length and SOC oscillation length are revealed.
 In Sec.~{\ref{numerical}}, both the
 analytical and numerical calculations for the 
steady-state spin diffusion in 3D isotropic speckle disorder are
presented. Specifically, the disorder
strength (Sec.~\ref{scattering_dependence}), Zeeman field strength
(Sec.~\ref{Zeeman_dependence})
 and SOC strength (Sec.~\ref{SOC_dependence}) dependencies are discussed. 
 We conclude and discuss in Sec.~{\ref{summary}}.

\section{Model and KSBEs}
\label{Model}
With the 3D disordered
speckle potential introduced to the spin diffusion
 region,\cite{Disorder_3D_Fermion,Disorder_3D_Boson,Disorder_1D_Boson,Disorder_theory}
 the Hamiltonian of the spin-orbit coupled ultracold atom,
 which is composed by the effective Hamiltonian $\hat{H}_0$,\cite{Fermi_SOC_1,Fermi_SOC_2} the disordered
speckle potential $U({\bf r})$,\cite{Disorder_3D_Fermion,Disorder_3D_Boson,Disorder_1D_Boson,
Disorder_theory,Configuration} 
  and the 
interatom interaction $\hat{H}_{\rm
  int}$, is written as 
\begin{equation}
\hat{H}=\hat{H}_0+U({\bf r})+\hat{H}_{\rm int}.
\end{equation}
The effective Hamiltonian consists of the kinetic energy of the atom and the SOC
($\hbar\equiv 1$),
\begin{equation}
H_0={\bf k}^2/(2m)+{\bgreek \Omega}({\bf k})\cdot {\bgreek \sigma}/2,
\label{Hamiltonian}
\end{equation}
with $\bgreek \sigma$ being the vector composed
of the Pauli matrices.
The interaction Hamiltonian $\hat{H}_{\rm int}$ is approximated by the s-wave interatom
scattering.\cite{interaction,interaction2,pwave1,Spielman,relaxation_1} In our
  study, the scattering length is tuned to be zero by the Feshbach
 resonance,\cite{Feshbach,Disorder_3D_Fermion,Disorder_3D_Boson,Disorder_1D_Boson,
Disorder_theory,Configuration} and hence $\hat{H}_{\rm int}$ is absent in our
following discussion.

The KSBEs, derived via the nonequilibrium Green function
method with the generalized Kadanoff-Baym
Ansatz,\cite{wu-review,2001,jianhua15,jianhua52,jianhua23}
 are utilized to study the spin diffusion in the ultracold Fermi gas:
\begin{equation}
  \partial_t \rho_{\bf k}({\bf r},t)=\partial_t\rho_{\bf k}({\bf r},t)|_{\rm dif}+
\partial_t\rho_{\bf k}({\bf r},t)|_{\rm coh}+\partial_t\rho_{\bf k}({\bf r},t)|_{\rm  scat}.
\label{ksbe}
\end{equation}
In these equations, $\rho_{\bf k}({\bf r},t)$ represent the density matrices of
atom with momentum ${\bf k}$ at position ${\bf r}=(x,y,z)$ and 
 time $t$, in which the diagonal elements $\rho_{{\bf k},\sigma\sigma}$ describe the atom distribution
 functions and the off-diagonal elements $\rho_{{\bf k},\sigma-\sigma}$ 
represent the correlation between the spin-up and down states.

For the quasi-one dimensional spin diffusion, the diffusion term is written as
\begin{equation}
\partial_t\rho_{\bf k}({\bf r},t)|_{\rm diff}=-(k_{\zeta}/m)
\partial_{\zeta} \rho_{\bf k}({\bf r},t),
\end{equation}
with $\zeta=x$ or $y$ for the spin diffusion along the $\hat{\bf x}$- or 
$\hat{\bf y}$-direction, respectively. 
 The coherent term is given by 
\begin{equation}
\partial_t\rho_{\bf k}({\bf r},t)|_{\rm
   coh}=-i\big[{\bf \Omega}({\bf
     k})\cdot{\bgreek\sigma}/2,\rho_{\bf
   k}({\bf r},t)\big],
\end{equation}
 where $[\ ,\ ]$ denotes the commutator.

 The scattering terms
 $\partial_t\rho_{\bf k}({\bf r},t)|_{\rm  scat}$ represent the atom-disorder
 scattering.\cite{Disorder_theory} 
 In our study, the effective Zeeman splitting energy and the 
SOC energy are much smaller than the Fermi energy.\cite{Jianhua} 
 Hence the atom-disorder scattering reads
\begin{equation}
\partial_t\rho_{\bf k}|^{ad}_{\rm  scat}=2\pi\sum_{{\bf k}'}|U_{{\bf k}-{\bf
    k}'}|^2\delta(\varepsilon_{{\bf k}}-\varepsilon_{{\bf k}'})(\rho_{{\bf
    k}'}-\rho_{{\bf k}}),
\label{a_disorder}
\end{equation}
where
\begin{eqnarray}
\nonumber
|U_{\bf q}|^2&=&\int\int d{\bf r}d{\bf r'}\langle[U({\bf
  r})-U_0] [U({\bf r'})-U_0]\rangle e^{-i{\bf q}\cdot({\bf
    r}-{\bf r'})}\\
&=&\int\int d{\bf r}d{\bf r'}C({\bf r}-{\bf r'})e^{-i{\bf q}\cdot({\bf
    r}-{\bf r'})}\equiv C_{\bf q}.
\label{UUq}
\end{eqnarray}
In Eq.~(\ref{UUq}), $U_0$ is the average value of the disorder
potential. For the 3D isotropic disordered
speckle,\cite{Disorder_3D_Fermion} 
\begin{equation}
C_{\bf q}=\pi^{3/2}V_R^2\sigma_R^3\exp(-\sigma_R^2q^2/4),
\end{equation}
where $V_R$ is the potential amplitude and $\sigma_R$ denotes the radius of
  the auto-correlation function of the laser.

In our model, with the same atom densities and hence the same chemical
  potentials in the left and right parts of the
  system, there is no ``charge'' diffusion in the system.
  The spin
  polarization at the boundary between the left and right parts of the system
  is approximately treated to be fixed.

\section{Analytical Results}
\label{analytical}
In this section, we analytically study the steady-state spin diffusion along the
  $\hat{\bf x}$- and $\hat{\bf y}$-directions in cold atoms with the atom-disorder
  scattering [Eq.~(\ref{a_disorder})]. Both the situations with spin
  polarization perpendicular (${\bf P}||\hat{\bf z}$) 
  and parallel (${\bf P}||\hat{\bf x}$) to the
  Zeeman field are analyzed.

In the steady state, the KSBEs are written as
\begin{eqnarray}
\nonumber
&&(k_{\zeta}/m)\partial_{\zeta} \rho_{{\bf k}}({\bf r})+i\Big[\Omega\sigma_x/2,\rho_{{\bf
    k}}({\bf r})\Big]+i\Big[\alpha k_x\sigma_z/2,\rho_{{\bf
    k}}({\bf r}) \Big]\\
&&\mbox{}+\sum\limits_{\bf k'}W_{\bf kk'}\big[\rho_{\bf k}({\bf r})-\rho_{\bf
  k'}({\bf r})\big]=0,
\label{KSBEs}
\end{eqnarray}
where $W_{\bf kk'}=2\pi C_{{\bf k}-{\bf k}'}\delta(\varepsilon_{\bf
  k}-\varepsilon_{{\bf k}'})$. 
In the strong scattering regime with $l_{\tau}\ll l_{\Omega},
l_{\alpha}$ [i.e., $\langle |\Omega({\bf k})|\rangle\tau_k\ll 1$],
 the steady-state diffusion lengths for different configurations 
are obtained and extended to the situation with moderate scattering strength ($l_{\Omega}\ll
l_{\tau}\ll l_{\alpha}$ or $l_{\alpha}\ll
l_{\tau}\ll l_{\Omega}$).

\subsection{Spin diffusion along the $\hat{\bf x}$-direction}
\label{analytical_x}
When the spin diffusion is along the
$\hat{\bf x}$-direction, by taking the steady-state condition, the Legendre
components of the azimuth-angle-averaged density matrix
[Eq.~(\ref{KSBEs_averaged})] with respect to the zenith angle $\theta_{\bf k}$
 are given by\begin{eqnarray}
\nonumber
\hspace{-0.6cm}&&\frac{k}{m}\frac{\partial}{\partial
  x}\Big[\frac{l\bar{\rho}_k^{l-1}}{\sqrt{2l-1}}+\frac{(l+1)\bar{\rho}_k^{l+1}}{\sqrt{2l+3}}\Big]
+i\Big[\frac{\Omega}{2}\sigma_x,\bar{\rho}_k^l\sqrt{2l+1}\Big]\\
\nonumber
\hspace{-0.6cm}&&\mbox{}+i\Big[\frac{\alpha
k}{2}\sigma_z,\frac{l\bar{\rho}_k^{l-1}}{\sqrt{2l-1}}+\frac{(l+1)\bar{\rho}_k^{l+1}}{\sqrt{2l+3}}\Big]
+\frac{\bar{\rho}_k^l}{\tau_{k,l}}\sqrt{2l+1}=0,\\
\hspace{-0.6cm}\label{KSBE_x}
\end{eqnarray}
in which the momentum relaxation time is denoted by
\begin{equation}
\tau_{k,l}^{-1}=\frac{m\sqrt{k}}{2\pi}\int_0^{\pi}C(\cos\theta)\big[1-P_l(\cos\theta)\big]\sin\theta d\theta,
\label{momentum_x}
\end{equation}
with $P_l(\cos\theta)$ being the Legendre function.
 By keeping both the zeroth and first orders
 ($l=0,1$), the analytical solution for the spin
 polarization is
obtained from Eq.~(\ref{KSBE_x}) (refer to Appendix~\ref{AA1}).  
It is found that for both the spin polarization 
  in the transverse ($\hat{\bf x}$-T) or longitudinal ($\hat{\bf x}$-L) configuration,
  the spin polarization is limited by one oscillation decay 
  together with one single exponential decay, i.e., 
\begin{equation}
{\bf S}^x_{\xi}\approx A_{\xi}\exp(-x/L_o^x)\cos(x/l_o^x)+B_{\xi}\exp(-x/L_s^x),
\label{single_oscillation_decay}
\end{equation}
with $\xi=t$ (transverse) or $l$ (longitudinal).
In Eq.~(\ref{single_oscillation_decay}), $A_{\xi}$ and $B_{\xi}$ are
 the amplitudes for the oscillation 
and single exponential decays, respectively, which are determined by the boundary condition;
$L_o^x$ and $L_s^x$ are the decay lengths for the oscillation and single
exponential decays,
respectively;
$l_o^x$ is the oscillation length for the oscillation decay. The integral forms
for $L_s^x$, $L_o^x$ and $l_o^x$ are complicated [Eqs.~(\ref{Lsx}),
  (\ref{Lox}) and (\ref{llox}) in
 Appendix~\ref{AA1}]. However, when the
system is in the strong ($l_{\tau}\ll l_{\Omega},
l_{\alpha}$) and moderate ($l_{\Omega}\ll
l_{\tau}\ll l_{\alpha}$ or $l_{\alpha}\ll
l_{\tau}\ll l_{\Omega}$) scattering regimes, it is found that the 
analytical results can be reduced to simple forms,
 which can describe the behavior of the spin diffusion quite well (Sec.~{\ref{numerical}}). 

 Specifically, for the oscillation decay, 
\begin{equation}
L_o^x\approx\left\{\begin{array}{cc}
\hspace{-0.5cm}2l_{\tau}/\sqrt{3},\hspace{1.1cm}{\rm when}\hspace{0.1cm}l_{\Omega}\ll l_{\tau}\ll l_{\alpha}\\
\hspace{-0.6cm}
\sqrt{2}l_{\tau}l_{\Omega}/(\sqrt{3}l_{\alpha}),\hspace{0.05cm}{\rm
  when}\hspace{0.1cm}l_{\alpha}\ll l_{\tau}\ll l_{\Omega}\\
\sqrt{2l_{\tau}l_{\Omega}/\sqrt{3}},\hspace{0.6cm}{\rm when}\hspace{0.1cm}l_{\tau}\ll l_{\alpha} \&l_{\tau}\ll l_{\Omega}
\end{array}\right..
\label{oscillation}
\end{equation}
The corresponding oscillation length   
\begin{equation}
l_o^x\approx\left\{\begin{array}{cc}
\hspace{-0.38cm}l_{\Omega},\hspace{1.45cm}{\rm when}\hspace{0.1cm}l_{\Omega}\ll l_{\tau}\ll l_{\alpha}\\
\hspace{-0.36cm} l_{\alpha},\hspace{1.42cm}{\rm when}\hspace{0.1cm}l_{\alpha}\ll
l_{\tau}\ll l_{\Omega}\\
\hspace{-0.08cm}\sqrt{2l_{\tau}l_{\Omega}/\sqrt{3}},
\hspace{0.25cm}{\rm when}\hspace{0.1cm}l_{\tau}\ll l_{\alpha} \&l_{\tau}\ll l_{\Omega}
\end{array}\right..
\label{oscillation_period}
\end{equation}
From Eq.~(\ref{oscillation_period}), one further notes that $l_{\Omega}$ and
 $l_{\alpha}$ correspond to the spacial oscillation
 length due the Zeeman and spin-orbit coupled fields. Because
 of this, we refer to $l_{\Omega}$
 and $l_{\alpha}$ as Zeeman and SOC oscillation lengths,
 respectively.
For the single exponential decay, the diffusion length reads
\begin{equation}
L_s^x\approx\left\{\begin{array}{cc}

\hspace{-0.5cm} l_{\tau}l_{\alpha}/(\sqrt{3}l_{\Omega}),\hspace{0.1cm}{\rm
  when}\hspace{0.1cm}l_{\Omega}\ll l_{\tau}\ll l_{\alpha}\\
\hspace{-0.5cm} l_{\tau}l_{\Omega}/(\sqrt{3}l_{\alpha}),\hspace{0.1cm}{\rm
  when}\hspace{0.1cm}l_{\alpha}\ll l_{\tau}\ll l_{\Omega}\\
\hspace{0.1cm}l_{\alpha},\hspace{1.6cm}{\rm when}\hspace{0.1cm} l_{\tau}\ll l_{\alpha} \&l_{\tau}\ll l_{\Omega}
\end{array}\right..
\label{single}
\end{equation}

From these simple dependencies of the spin diffusion length on the mean free
  path, the Zeeman oscillation length and the SOC oscillation length, the
  behavior of the steady-state spin polarization shown
 in Eq.~(\ref{single_oscillation_decay}) can be further
  simplified.
 It can be demonstrated 
  that when $L_o^x\approx L_s^x$, $A_{\xi}\approx B_{\xi}$. Specifically, only
  when $L_o^x\approx L_s^x$, both the oscillation decay and single exponential
  decay are
  important, but with similar decay length. Otherwise, the spin polarization
 can be approximately reduced to one oscillation or single exponential decay,
which depends on its amplitude in the spin polarization. In different regimes,
 the different behaviors
for the steady-state spin polarization 
 are analyzed as follows. In the Zeeman-field (SOC) dominated moderate scattering regime 
 with $l_{\Omega}\ll l_{\tau}\ll l_{\alpha}$ ($l_{\alpha}\ll l_{\tau}\ll
  l_{\Omega}$),
the condition $L_o^x\approx L_s^x$ is never satisfied. Accordingly,
 when $l_{\Omega}\ll l_{\tau}\ll l_{\alpha}$ ($l_{\alpha}\ll l_{\tau}\ll
  l_{\Omega}$), in the
transverse/longitudinal  situation, the
steady-state 
spin polarization is approximately oscillation/single exponential
(single exponential/oscillation)
decay.
 In the strong scattering regime ($l_{\tau}\ll l_{\alpha}, l_{\Omega}$),
 when $L_o^x\approx L_s^x$, one obtains 
\begin{equation}
l_{\tau,x}^c\approx\sqrt{3}l_{\alpha}^2/(2l_{\Omega}),
\label{critical}
\end{equation}
which is referred to as the crossover length between the
  relatively strong and strong scattering regimes.
Accordingly, when
$l_{\tau}\gg l_{\tau,x}^c$ ($l_{\tau}\ll l_{\tau,x}^c$), the steady-state spin polarization is
approximated by a single exponential/oscillation
(oscillation/single exponential) decay in the
transverse/longitudinal situation. 

Based on the above analysis, we summarize the behaviors of the steady-state
  spin polarization and the spin diffusion lengths in Table~\ref{table_x} for the two specific configurations
$\hat{\bf x}$-T and $\hat{\bf x}$-L in
different regimes.
 As shown in the table, 
the system is divided
   into five regimes: I, weak scattering regime ($l_{\tau}\gg l_{\Omega},
    l_{\alpha}$); II, Zeeman field-dominated moderate scattering regime
 ($l_{\Omega}\ll l_{\tau}\ll l_{\alpha}$); III, SOC-dominated moderate
 scattering regime ($l_{\alpha}\ll l_{\tau}\ll
    l_{\Omega}$); IV, relatively strong scattering regime
    ($l_{\tau,x}^c\ll l_{\tau}\ll l_{\alpha}, l_{\Omega}$); V, strong
    scattering regime ($l_{\tau}\ll l_{\alpha}, l_{\Omega},l_{\tau,x}^c$).
 In different regimes, it can be
seen that the dependencies of the spin diffusion length on the scattering
  strength, the Zeeman field and SOC strength
 are very rich. Specifically, in the Zeeman field- (SOC-) dominated
moderate scattering regime with $l_{\Omega}\ll l_{\tau}\ll l_{\alpha}$
 ($l_{\alpha}\ll l_{\tau}\ll l_{\Omega}$),  
the longitudinal (longitudinal/transverse) spin diffusion is enhanced by the Zeeman
field (SOC);
whereas in the relatively strong
 (strong) scattering regime ($l_{\tau}\ll l_{\alpha}, l_{\Omega}$), 
the transverse (longitudinal) spin diffusion is determined by only the SOC
oscillation length, but irrelevant to the Zeeman
field with
$l_{\tau}\gg l_{\tau,x}^c$ ($l_{\tau}\ll l_{\tau,x}^c$). 
The rich behaviors of the spin diffusions in different regimes
 are hard to be understood in the framework of the previous simple drift-diffusion
  model\cite{Peng43,Peng44,Peng45,Peng46,Peng47,Peng} or the direct inhomogeneous
  broadening [Eqs.~(\ref{diffusion_kx}) and (\ref{diffusion_ky})]
 picture\cite{2001,wu-review,diffusion_k,double,JinLuo_infinite}
 in the literature. In the simple drift-diffusion model, there are only
  two rather than {\em five} 
  regimes: the strong
 scattering regime with $l_s\propto 1/(m|\alpha|)$ and the weak scattering
 regime with $l_s\propto k\sqrt{\tau_k}/m$. In the direct inhomogeneous
broadening picture, from Eq.~(\ref{diffusion_kx}),
 the Zeeman field (SOC) can (cannot) provide the inhomogeneous
broadening. Hence, it seems that the spin diffusion can be suppressed by the
Zeeman field, but irrelevant to the SOC. Below, we extend our previous inhomogeneous
  broadenings [Eqs.~(\ref{diffusion_kx}) and (\ref{diffusion_ky})] to the
  effective ones. We will show that based on the {\it effective}
 inhomogeneous broadening, apart from Regime
IV, the above anomalous behaviors of the spin
diffusion can be understood from the view point of the helix
 representation.\cite{relaxation_3,Anomalous_DP_2} In the helix
   space, apart from the spin-conserving scattering [the fifth term in the left-hand side
of Eq.~(A6) in our previous work\cite{relaxation_3}], 
additional terms arise including the helix spin-flip scattering [the sixth term in the left-hand side
of Eq.~(A6) in Ref.~\onlinecite{relaxation_3}] and helix coherence term [the last term
in the left-hand side of Eq.~(A6) in Ref.~\onlinecite{relaxation_3}].\cite{relaxation_3,Anomalous_DP_2}

\begin{widetext}
\begin{center}
\begin{table}[htb]
  \caption{Behaviors of the steady-state
  spin polarization in the spatial domain and corresponding spin diffusion lengths for configurations
$\hat{\bf x}$-T and $\hat{\bf x}$-L in
different regimes.}
  \label{table_x} 
  \begin{tabular}{l| l| l |l}
    \hline
    \hline
    \;\;\;\;\;\;\;\;\;\;\;\;\;\;\;Regime&\;\;\;\;\;\;\;\;\;\;\;Condition&\;\;\;\;\;Behavior and $L_T^x$ in $\hat{\bf
      x}$-T\;\;&\;\;\;\;\;Behavior and $L_{L}^x$ in $\hat{\bf x}$-L\;\;\\
    \hline  
    I: weak scattering regime&\;\;\;\;\;\;\;\;\;$l_{\tau}\gg l_{\Omega},
    l_{\alpha}$&\hspace{0.8cm}\;\;\;\;\;\;\;\;\;\;\;NA&\hspace{0.8cm}\;\;\;\;\;\;\;\;\;\;\;\;NA\\
\hline
    II: Zeeman field-dominated&\;\;\;\;\;\;\;\;$l_{\Omega}\ll l_{\tau}\ll l_{\alpha}$&\hspace{0.8cm}\;\;\;oscillation
     decay;&\hspace{0.1cm}\;\;\;\;\;single exponential decay;\;\;\;\;\\
    moderate scattering
    regime&&\hspace{0.8cm}\;\;\;\;\;\;\;$2l_{\tau}/\sqrt{3}$&\hspace{0.8cm}\;\;\;\;\;\;$l_{\tau}l_{\alpha}/(\sqrt{3}l_{\Omega})$\\
\hline
    III: SOC-dominated moderate&\;\;\;\;\;\;\;$l_{\alpha}\ll l_{\tau}\ll
    l_{\Omega}$&
\hspace{0.1cm}\;\;\;\;\;single exponential decay;\;\;\;\;&\hspace{0.8cm}\;\;\;oscillation
    decay;\;\;\;\;\\
    scattering
    regime&&\hspace{0.95cm}\;\;\;\;\;$l_{\tau}l_{\Omega}/(\sqrt{3}l_{\alpha})$&
\hspace{0.95cm}\;\;$\sqrt{2}l_{\tau}l_{\Omega}/(\sqrt{3}l_{\alpha})$\\
\hline
    IV: relatively strong scattering&\;\;\;\;$l_{\tau}\ll l_{\alpha}, l_{\Omega} \&l_{\tau}\gg
    l_{\tau,x}^c$\;\;\;\;&\hspace{0.1cm}\;\;\;\;\;single exponential
    decay;\;\;\;\;&\hspace{0.8cm}\;\;\;oscillation decay;\;\;\;\;\\
      regime&\;\;\;\;\;\;\;\;\;($l_{\tau}\ll l_{\alpha}\ll l_{\Omega}$)\;\;\;\;
 &\hspace{0.9cm}\;\;\;\;\;\;\;\;\;\;\;\;$l_{\alpha}$&\hspace{0.8cm}\;\;\;\;\;$\sqrt{2l_{\tau}l_{\Omega}/\sqrt{3}}$\\
\hline
 V: strong scattering regime&\;\;\;\;$l_{\tau}\ll l_{\alpha}, l_{\Omega} \&l_{\tau}\ll
    l_{\tau,x}^c$\;\;\;\;&\hspace{0.8cm}\;\;\;oscillation
    decay;&\hspace{0.1cm}\;\;\;\;\;single exponential 
    decay;\;\;\;\;\\
    &&\hspace{0.8cm}\;\;\;\;\;
$\sqrt{2l_{\tau}l_{\Omega}/\sqrt{3}}$&\hspace{0.9cm}\;\;\;\;\;\;\;\;\;\;\;\;$l_{\alpha}$\\
    \hline
    \hline
\end{tabular}\\
\hspace{-14.5cm}$l_{\tau,x}^c\approx\sqrt{3}l_{\alpha}^2/(2l_{\Omega})$. 
\end{table}
\end{center}
\end{widetext}

\begin{figure}[ht]
  {\includegraphics[width=7.5cm]{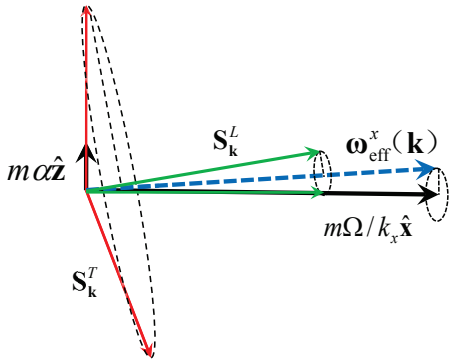}}
  \caption{(Color online) Schematic for the spin precession around the
    effective inhomogeneous broadening field ${\bgreek \omega}^x_{\rm eff}({\bf
  k})$ [Eq.~(\ref{effective_Omega})] in the transverse (${\bf S}_{\bf k}^T$)
 and longitudinal (${\bf S}_{\bf k}^L$)
spin diffusions. With the strong Zeeman and weak spin-orbit coupled fields, for
the transverse (longitudinal) situation, the spin vector ${\bf S}_{\bf k}^T$
 (${\bf S}_{\bf k}^L$) is
perpendicular (parallel) to ${\bgreek \omega}^x_{\rm eff}({\bf
  k})$ approximately. Therefore, the inhomogeneous broadening field can (cannot)
cause efficient
spin precession in the transverse (longitudinal) spin diffusion.}
  \label{figyw1}
\end{figure}

When the system is in Regime II, the Zeeman
    field-dominated moderate scattering regime ($l_{\Omega}\ll l_{\tau}\ll
    l_{\alpha}$) or Regime V, the strong scattering regime
 ($l_{\tau}\ll l_{\alpha}, l_{\Omega}\ll l_{\tau,x}^c$), the condition $l_{\Omega}\ll
l_{\alpha}$ is satisfied. Therefore, both the transverse and longitudinal spin
diffusions can be understood in the limit with strong Zeeman and weak spin-orbit coupled
fields. In this situation, the effective inhomogeneous broadening field is given by
\begin{eqnarray} 
\nonumber
{\bgreek \omega}^x_{\rm eff}({\bf k})&=&(m/k_x)\sqrt{\Omega^2 +\alpha^2
  k_x^2}\hat{\bf x}'\\
&\approx&
\big[m\Omega/k_x+m\alpha^2
  k_x/(2\Omega)\big]\hat{\bf x}',
\label{effective_Omega}
\end{eqnarray}
with $\hat{\bf x}'=\frac{\displaystyle 1}{\displaystyle \sqrt{1+(\alpha
      k_x/\Omega)^2}}\hat{\bf x}+\frac{\displaystyle \alpha
  k_x/\Omega}{\displaystyle \sqrt{1+(\alpha k_x/\Omega)^2}} \hat{\bf z}$ being nearly parallel
  to $\hat{\bf x}$. Under this effective field, the behaviors of 
  the spin precession in the spacial domain for the transverse and longitudinal
  spin diffusions can be obtained, as schematically shown in Fig.~\ref{figyw1}.  
In this figure, for the transverse spin diffusion, the spin polarization is
perpendicular to the strong Zeeman field and hence ${\bgreek \omega}^x_{\rm eff}({\bf
  k})$ approximately. During the scattering, the spin vectors rotate around
the effective inhomogeneous broadening field ${\bgreek \omega}^x_{\rm eff}({\bf
  k})$
 fast. Moreover, the scattering can also influence the spin
diffusion. In the helix space, as mentioned above, apart from the original
spin-conserving scattering, there exist the helix spin-flip scattering and helix
coherence processes. However, with the helix spin-flip rate 
  $\alpha^2k^2/(\Omega^2\tau_{k})\ll 1/\tau_k$ and helix coherence rate
 $\alpha k/(\Omega\tau_{k})\ll 1/\tau_k$ when $\Omega\gg \alpha k$, both the helix
spin-flip scattering and helix coherence can be neglected. In this situation,
 the effective inhomogeneous broadening together with the spin-conserving scattering determines the
behavior of the spin diffusion. We refer to this picture as the {\it modified}
  inhomogenous broadening picture. For the longitudinal situation, the spin polarization is nearly
parallel to ${\bgreek \omega}^x_{\rm eff}({\bf
  k})$, and hence the effective inhomogeneous
broadening cannot cause the spin precession effectively. In this situation,
 the spin diffusion can be understood from the 
drift-diffusion model\cite{Peng43,Peng44,Peng45,Peng46,Peng47,Peng} modified as follows.
 The diffusion length $l_s=\sqrt{D\tau_s({\bf k})}$
 in which  
$D=v_F^2\tau_{k}/3$ is the diffusion
coefficient with $v_F$ being the Fermi velocity. 
The SRT is analyzed in the helix space.
 First of all, for the longitudinal situation here, the helix coherence term has
  no contribution to the spin relaxation. In this situation, there exit two
 channels influencing the spin relaxation:
 (i), the effective inhomogenous broadening together with the spin-conserving
  scattering; (ii),
  the helix spin-flip scattering.\cite{relaxation_3,Anomalous_DP_2} In the
  strong scattering regime, both channels (i) and (ii) are important for the
  spin relaxation, in which $\tau_s({\bf k})$ remains the SRT in the
   conventional DP
   mechanism.\cite{relaxation_1,relaxation_2,relaxation_4,wu-review} In the
   moderate scattering regime, channel (ii) is dominant for the spin relaxation,
 and hence $\tau_s({\bf k})$ is replaced by
 the helix spin-flip rates determined in the helix
  space.\cite{relaxation_3,Anomalous_DP_2} We refer to the above pictures as {\it modified} drift-diffusion
  model. 
  Accordingly, in the moderate scattering regime, the dependence of the 
 helix spin-flip rate on the scattering strength, the Zeeman
field and SOC strength disclosed in our
previous works\cite{relaxation_3,Anomalous_DP_2,Anomalous_DP_1} can also 
influence the spin diffusion.

Specifically, in the Zeeman field-dominated moderate scattering regime ($l_{\Omega}\ll
 l_{\tau}\ll l_{\alpha}$), i.e., Regime II, the Zeeman oscillation length is
 the shortest length scale in the spin diffusion, which determines the behavior of spin
 precession in the spacial domain during the scattering. When the
spin polarization is perpendicular to the Zeeman field (transverse
configuration), the spin vector approximately precesses around $m\Omega/k_x
\hat{\bf x'}$, leading to the spacial oscillations with the 
period proportional to $\langle |k_x|\rangle/(m\Omega)$
 [$l_o^x\approx l_{\Omega}$ in Eq.~(\ref{oscillation_period})].
 Moreover, due to the fast spacial oscillations with the strong Zeeman field,
 the spin memory is
lost during one spin-conserving scattering, with the diffusion length being approximately the mean
free path ($L_{T}^x\approx 2l_{\tau}/\sqrt{3}$ in Table~\ref{table_x}).
 When the spin polarization is parallel to the Zeeman field (longitudinal
  configuration), the effective inhomogeneous broadening ${\bgreek \omega}^x_{\rm eff}({\bf
    k})$ cannot cause spin precession efficiently and the steady-state spin polarization 
decays without any oscillation. The spin diffusion can be understood from
the modified drift-diffusion model.\cite{Peng43,Peng44,Peng45,Peng46,Peng47,Peng}
Specifically, the helix spin-flip rate is calculated to be
$\alpha^2k^2/(2\Omega^2\tau_k)$
  in the moderate scattering
 situation.\cite{relaxation_3,Anomalous_DP_2,Anomalous_DP_1} Accordingly, the
 spin diffusion length in the modified drift-diffusion model is given
 by $l_s\approx \sqrt{2}l_{\tau}l_{\alpha}/(\sqrt{3}l_{\Omega})$, which is consistent
 with our model shown as $L_L^x\approx l_{\tau}l_{\alpha}/(\sqrt{3}l_{\Omega})$
 in Table~\ref{table_x}. Specifically, one notes that due to the suppression of
 the spin relaxation by the Zeeman field, the longitudinal spin diffusion length is enhanced by the
 Zeeman field.

In Regime V, the strong scattering regime ($l_{\tau}\ll
  l_{\alpha}, l_{\Omega}, l_{\tau,x}^c$), we consider a limit situation with the
Zeeman field much stronger than the spin-orbit coupled one ($l_{\tau}\ll
  l_{\alpha}, l_{\Omega}\ll l_{\tau,x}^c$). The
effective inhomogeneous broadening is given by Eq.~({\ref{effective_Omega}}). For the transverse
spin diffusion, because the inhomogeneous
broadening given by the Zeeman field is dominant, the spin diffusion length is suppressed
by the Zeeman field, but less influenced by the SOC. Moreover, 
 during the diffusion, the atoms experience several
  spin-conserving scatterings, which suppress
  the spin diffusion. 
 Accordingly, we obtain a reasonable picture to understand
 $L_{L}^x\approx \sqrt{2l_{\tau}l_{\Omega}/\sqrt{3}}$ in
Table~\ref{table_x}.
  For the longitudinal spin diffusion, the inhomogeneous
broadening cannot cause the spin precession efficiently, with the steady-state spin
polarization showing single exponential
decay. Hence, the modified
drift-diffusion model can be used.\cite{Peng43,Peng44,Peng45,Peng46,Peng47,Peng}
 With the SRT in the strong scattering regime $\tau_{s}({\bf
  k})\approx 2/(\alpha^2k^2\tau_k)$,\cite{relaxation_3,Anomalous_DP_2,Anomalous_DP_1}
   one obtains the spin diffusion length 
$l_s\approx \sqrt{2}l_{\alpha}/\sqrt{3}$ ($L_T^x\approx l_{\alpha}$ in
Table~\ref{table_x}).
Therefore, the longitudinal spin diffusion length depends only on the SOC oscillation
length.\cite{relaxation_3,Peng43,Peng44,Peng45,Peng46,Peng47,Peng}

\begin{figure}[ht]
  {\includegraphics[width=7.5cm]{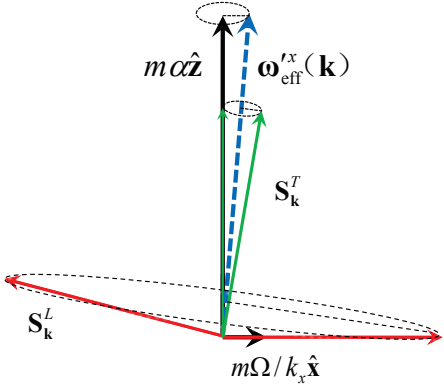}}
  \caption{(Color online) Schematic for the spin precession around the
    effective inhomogeneous broadening field ${\bgreek \omega'}^x_{\rm eff}({\bf
  k})$ [Eq.~(\ref{effective_alpha})] in the transverse (${\bf S}_{\bf k}^T$)
 and longitudinal (${\bf S}_{\bf k}^L$)
spin diffusions. With the weak Zeeman and strong spin-orbit coupled fields, for
the transverse (longitudinal) situation, the spin vector ${\bf S}_{\bf k}^T$
 (${\bf S}_{\bf k}^L$) is
parallel (perpendicular) to ${\bgreek \omega'}^x_{\rm eff}({\bf
  k})$ approximately. Therefore, the inhomogeneous broadening field cannot (can)
cause efficient
spin precession in the transverse (longitudinal) spin diffusion.}
  \label{figyw2}
\end{figure}

When the system lies in Regime III, the SOC-dominated moderate scattering
  regime
 ($l_{\alpha}\ll l_{\tau}\ll l_{\Omega}$) and Regime IV, the
relatively strong scattering regime
 ($l_{\tau,x}^c\ll l_{\tau}\ll l_{\alpha}, l_{\Omega}$), the spin-orbit coupled
 field is much stronger than the Zeeman one. In this situation,
 the effective inhomogeneous broadening field reads
\begin{eqnarray} 
\nonumber
{\bgreek \omega'}^x_{\rm eff}({\bf k})&=&(m/k_x)\sqrt{\alpha^2
  k_x^2+\Omega^2}\hat{\bf z}'\\
&\approx&
\big[m\alpha+m\Omega^2/(2\alpha k_x^2)\big]\hat{\bf z}',
\label{effective_alpha}
\end{eqnarray}
where $\hat{\bf z}'=\frac{\displaystyle 1}{\displaystyle \sqrt{1+[\Omega/(\alpha k_x)]^2}}\hat{\bf
  z}+\frac{\displaystyle \Omega/(\alpha k_x)}{\displaystyle
  \sqrt{1+[\Omega/(\alpha k_x)]^2}}\hat{\bf x}$
 is parallel to $\hat{\bf z}$ approximately. The analysis is similar to the
   situation with 
 strong Zeeman and weak spin-orbit coupled fields. One obtains that for the
 transverse (longitudinal) spin diffusion, the spin polarization is nearly 
parallel (perpendicular) to ${\bgreek \omega'}^x_{\rm
  eff}({\bf k})$, which cannot (can) cause efficient spin precession.
  These pictures are summarized in Fig.~\ref{figyw2}.  
Therefore, 
 the behavior of the transverse spin diffusion can be
understood from the modified drift-diffusion
model;\cite{Peng43,Peng44,Peng45,Peng46,Peng47,Peng} whereas 
the longitudinal spin diffusion can be analyzed from 
 the features of the precession of the spin vectors around ${\bgreek \omega'}^x_{\rm
  eff}({\bf k})$.

Specifically, in Regime III, i.e., the SOC-dominated moderate scattering regime ($l_{\alpha}\ll
 l_{\tau}\ll l_{\Omega}$), in the
 transverse configuration, the modified drift-diffusion model is used to
 understand the spin
 diffusion.\cite{Peng43,Peng44,Peng45,Peng46,Peng47,Peng} In the helix
 representation, the helix spin-flip rate is proportional
 to $\Omega^2/(\alpha^2 k^2\tau_k)$ approximately.\cite{relaxation_3,Anomalous_DP_2,Anomalous_DP_1}
 The corresponding spin diffusion length is
 proportional to $l_{\tau}l_{\Omega}/l_{\alpha}$, which is consistent with $L_T^x\approx
 l_{\tau}l_{\Omega}/(\sqrt{3}l_{\alpha})$ in Table~\ref{table_x}. Consequently,
 one observes that the spin relaxation is suppressed by the
 spin-orbit coupled field, and the transverse spin diffusion length is enhanced by
 the SOC. In the longitudinal configuration, with
$m|\alpha|\gg m\Omega^2/(2|\alpha| k_x^2)$, the spin
polarization evolves with oscillations in the spacial domain, whose oscillation
length is $l_o^x\approx l_{\alpha}$ [Eq.~(\ref{oscillation_period})].
 Furthermore, only $m\Omega^2/(2\alpha k_x^2){\hat{\bf z'}}$ in
${\bgreek \omega'}^x_{\rm eff}({\bf k})$ can cause the inhomogeneous broadening.
 Therefore, it can be expected that
the longitudinal 
spin diffusion length is inversely proportional (proportional) to the Zeeman
field (SOC).  
Moreover, the spin-conserving scattering can suppress the spin diffusion in
  the modified inhomogeneous broadening picture. These
analysis are consistent with $L_T^x\approx
\sqrt{2}l_{\tau}l_{\Omega}/(\sqrt{3}l_{\alpha})$ in Table~\ref{table_x}.

In Regime IV, the relatively strong scattering regime ($l_{\tau,x}^c\ll l_{\tau}\ll l_{\alpha}, l_{\Omega}$),
  for the transverse spin diffusion, from the modified drift-diffusion
    model,\cite{Peng43,Peng44,Peng45,Peng46,Peng47,Peng} $l_s\approx
    \sqrt{2}l_{\alpha}/\sqrt{3}$ ($L_T^x\approx l_{\alpha}$ in
Table~\ref{table_x}).
 For the longitudinal spin diffusion, one expects that the modified
   inhomogeneous broadening picture can be applied. However, this
picture fails to explain the behavior of the longitudinal spin diffusion.
It can be seen from Eq.~(\ref{effective_alpha}) that both the spin-orbit coupled and Zeeman
 fields
 provide the inhomogeneous broadening [$m\Omega^2/(2\alpha
 k_x^2){\hat{\bf z'}}$ in Eq.~(\ref{effective_alpha})]. By further
   considering that the spin diffusion is suppressed by the spin-conserving
   scattering, it is obtained that the spin diffusion length is proportional to
   the SOC strength, but inversely proportional to the Zeeman field strength and
   scattering. However, in Table~\ref{table_x}, the longitudinal spin diffusion
   length $L_{L}^x\approx \sqrt{2l_{\tau}l_{\Omega}/\sqrt{3}}$ is irrelevant to
   the SOC strength. One further notices that Regime IV lies in the crossover
   region 
   between the moderate and strong scattering regimes. When the scattering is relatively
strong, the shortest length scale in the spin diffusion is the mean free
path. However, there still exists strong competition
 between the effective inhomogeneous broadening and scattering, which makes the
 behavior of the spin diffusion complicated.\cite{Peng,JinLuo_infinite,double} 

Finally, we address the behavior of the spin diffusion along the $\hat{\bf
  x}$-direction
 in the limit situation where the Zeeman field is zero.
 When $\Omega=0$, $l_{\Omega}$ is infinite. The corresponding steady-state spin
polarization shows very different behaviors compared to the situation with
finite Zeeman field. In this situation, the transverse (longitudinal)
spin diffusion length is infinite because ${\bgreek \omega'}^x_{\rm eff}({\bf
  k})$ cannot cause inhomogeneous broadening.\cite{JinLuo_infinite}
Specifically, for the longitudinal situation, the spin polarization is perpendicular to
the spin-orbit coupled field, spin helix establishes with the oscillation length
being $l_{\alpha}$.\cite{relaxation_1,JinLuo_infinite}

\subsection{Spin diffusion along the $\hat{\bf y}$-direction}
When the spin diffusion is along the
$\hat{\bf y}$-direction,
 by taking the steady-state condition, the Fourier
components of the azimuth-angle-averaged density matrix
[Eq.~(\ref{KSBEs_averaged})] with respect to the zenith angle $\theta_{\bf k}$
 are given by
\begin{eqnarray}
\nonumber
&&\frac{\hbar k}{2im}\frac{\partial}{\partial
  y}\big(\tilde{\rho}_k^{l-1}-\tilde{\rho}_k^{l+1}\big)
+i\Big[\frac{\alpha
k}{4}\sigma_z,\tilde{\rho}_k^{l+1}+\tilde{\rho}_k^{l-1}\Big]\\
&&\mbox{}+i\Big[\frac{\Omega}{2}\sigma_x,\tilde{\rho}_k^l\Big]
+\frac{\tilde{\rho}_k^l}{\tilde{\tau}_{k,l}}=0,
\label{KSBE_y}
\end{eqnarray}
in which the momentum relaxation time is denoted by
\begin{equation}
\tilde{\tau}_{k,l}^{-1}=\frac{m\sqrt{k}}{2\pi}\int_0^{\pi}C(\cos\theta)\big[1-\cos(l\theta)\big]\sin\theta d\theta.
\label{momentum_y}
\end{equation}
One notes that by choosing the Legendre and Fourier expansions of the density
matrix for the spin
diffusions along the $\hat{\bf x}$- and $\hat{\bf y}$-directions, the definitions for the
momentum scattering time $\tau_{k,l}$ [Eq.~(\ref{momentum_x})] and
$\tilde{\tau}_{k,l}$ [Eq.~(\ref{momentum_y})] are different. However, when
$l=1$, the two definitions are the same. By keeping both the zeroth and first
orders ($l=0,1$), the
analytical solutions for both the spin polarization perpendicular ($\hat{\bf y}$-T) and
parallel ($\hat{\bf y}$-L) to the Zeeman field are obtained (refer to Appendix~\ref{AA2}).

When the spin polarization is perpendicular to the Zeeman field, in the
  moderate and strong scattering regimes, similar to the spin diffusion along
  the $\hat{\bf
    x}$-direction (Table~{\ref{table_x}}),
 the behaviors of the steady-state spin polarization are 
  different with different scattering strengths. When the scattering is
  relatively weak, which satisfies $l_{\tau}>l_{\tau,y}^c$ with $l_{\tau,y}^c=\frac{\displaystyle
    4l_{\alpha}}{\displaystyle \sqrt{3}l_{\Omega}}\Big(\frac{\displaystyle
      1}{\displaystyle l_{\alpha}^2}-\frac{\displaystyle 8}{\displaystyle
      3l_{\Omega}^2}\Big)^{-1/2}$, the steady-state spin polarization for ${{\bf S}}^y_T$
  is limited by the bi-exponential decay, i.e.,  
\begin{equation}
{\bf S}^y_T=P^+\exp(-y/L^{y,+}_T)+P^-\exp(-y/L^{y,-}_T),
\label{bi_exponential_y}
\end{equation}
with $L^{y,\pm}_T$ being the diffusion length.
 It is further demonstrated that when
$L^{y,+}_T\ll L^{y,-}_T$, $P^+\ll P^-$ and hence the spin
polarization [Eq.~(\ref{bi_exponential_y})] reduces to a single exponential decay with the decay length being
$L^{y,-}_T$. Specifically, in the moderate scattering regime 
with $l_{\alpha}\ll l_{\tau}\ll l_{\Omega}$ ($l_{\Omega}\ll l_{\tau}\ll
l_{\alpha}$), the condition $l_{\tau}>l_{\tau,y}^c$ is naturally (never) satisfied; 
in the strong scattering regime ($l_{\tau,y}^c<l_{\tau}\ll
l_{\alpha}, l_{\Omega}$), it can be obtained that $l_{\alpha}\ll l_{\Omega}$ and hence $l_{\tau,y}^c\approx
4l_{\alpha}^2/(\sqrt{3}l_{\Omega})$. Accordingly, when $l_{\tau}\gg
  l_{\tau,y}^c$,
 the diffusion length for ${\bf S}^y_T$ is
written as
\begin{equation}
L^{y,-}_T\approx\left\{\begin{array}{cc}
\hspace{-0.3cm} {\rm NA},\hspace{1.3cm}{\rm
  when}\hspace{0.1cm}l_{\Omega}\ll l_{\tau}\ll l_{\alpha}\\
\hspace{-0.7cm} \sqrt{3}l_{\tau}l_{\Omega}/(2l_{\alpha}), \hspace{0.3cm}{\rm
  when}\hspace{0.1cm}l_{\alpha}\ll l_{\tau}\ll l_{\Omega}\\
\hspace{0cm}\sqrt{3}l_{\tau}l_{\Omega}/(2l_{\alpha}),\hspace{0.3cm}{\rm when}\hspace{0.1cm} l_{\tau,y}^c\ll l_{\tau}\ll
l_{\alpha}, l_{\Omega}
\end{array}\right..
\label{single_ST}
\end{equation}

When the scattering is relatively strong, which satisfies
 $l_{\tau}<l_{\tau,y}^c$, 
  the transverse spin polarization in the steady state is determined by the oscillation decay, i.e.,  
\begin{equation}
    {\bf S}^y_T=P_0\exp(-y/L^y_{T})\cos(y/l^y_{T}).
  \end{equation}
Here, the decay length $L^y_{T}$ and oscillation length $l^y_{T}$ can be written as
\begin{equation}
L^y_{T}\approx\left\{\begin{array}{cc}
\hspace{-0.25cm} \sqrt{2}l_{\tau},\hspace{1cm}{\rm
  when}\hspace{0.1cm}l_{\Omega}\ll l_{\tau}\ll l_{\alpha}\\
\hspace{0.05cm} {\rm NA},\hspace{0.95cm}{\rm
  when}\hspace{0.1cm}l_{\alpha}\ll l_{\tau}\ll l_{\Omega}\\
\hspace{-0.15cm}\sqrt{\sqrt{3}l_{\tau}l_{\Omega}},\hspace{0.6cm}{\rm
  when}\hspace{0.1cm}
 l_{\tau}\ll l_{\alpha}, l_{\Omega}, l_{\tau,y}^c
\end{array}\right.
\label{oscillation_Sz}
\end{equation}
and 
\begin{equation}
l^y_{T}\approx\left\{\begin{array}{cc}
\hspace{-0.15cm} \sqrt{3/2}l_{\Omega},\hspace{0.5cm}{\rm
  when}\hspace{0.1cm}l_{\Omega}\ll l_{\tau}\ll l_{\alpha}\\
\hspace{0.2cm} {\rm NA},\hspace{0.85cm}{\rm
  when}\hspace{0.1cm}l_{\alpha}\ll l_{\tau}\ll l_{\Omega}\\
\hspace{-0.15cm}\sqrt{\sqrt{3}l_{\tau}l_{\Omega}},\hspace{0.6cm}{\rm
  when}\hspace{0.1cm}
 l_{\tau}\ll l_{\alpha}, l_{\Omega}, l_{\tau,y}^c
\end{array}\right.,
\label{oscillation_Sz_period}
\end{equation}
respectively.

When the spin polarization is parallel to the Zeeman field, it is found
  that the steady-state spin
  polarization is limited by the single exponential decay, i.e.,
\begin{equation}
{\bf S}^y_L=P_0\exp(-y/L^y_{L}).
\end{equation}
Here, $L^y_{L}=l_{\alpha}\sqrt{l_{\tau}^2/(3l_{\Omega}^2)+1}$ is the decay
length for ${{\bf S}}^y_L$. In the moderate and strong scattering regimes,
\begin{equation}
L^y_{L}\approx\left\{\begin{array}{cc}
\hspace{-0.7cm} l_{\tau}l_{\alpha}/(\sqrt{3}l_{\Omega}),\hspace{0.1cm}{\rm
  when}\hspace{0.1cm}l_{\Omega}\ll l_{\tau}\ll l_{\alpha}\\
\hspace{-0.3cm} l_{\alpha}, \hspace{1.27cm}{\rm
  when}\hspace{0.1cm}l_{\alpha}\ll l_{\tau}\ll l_{\Omega}\\
\hspace{0.3cm}l_{\alpha},\hspace{1.3cm}{\rm when}\hspace{0.1cm} l_{\tau}\ll l_{\Omega}\&l_{\tau}\ll l_{\alpha}
\end{array}\right..
\label{oscillation_Sx}
\end{equation}

Based on the above results, in
different regimes, the behaviors of the steady-state
  spin polarization and diffusion lengths 
are summarized in Table~\ref{table_y} for the two specific configurations
$\hat{\bf y}$-T and $\hat{\bf y}$-L.
\begin{widetext}
\begin{center}
\begin{table}[htb]
  \caption{Behaviors of the steady-state
  spin polarization in the spacial domain and corresponding spin diffusion lengths for configurations
$\hat{\bf y}$-T and $\hat{\bf y}$-L in
different regimes.}
  \label{table_y} 
  \begin{tabular}{l| l| l |l}
    \hline
    \hline
    \;\;\;\;\;\;\;\;\;\;\;\;\;\;\;Regime&\hspace{0.2cm}\;\;\;\;\;\;\;\;\;\;\;\;Condition&\;\;\;\;\;Behavior
    and $L_T^y$ in $\hat{\bf
      y}$-T\;\;&\;\;\;\;\;Behavior and $L_{L}^y$ in $\hat{\bf y}$-L\;\;\\
    \hline  
    I: weak scattering regime&\hspace{0.75cm}$l_{\tau}\gg  l_{\Omega},
    l_{\alpha}$&\hspace{0.8cm}\;\;\;\;\;\;\;\;\;\;\;NA&\hspace{0.8cm}\;\;\;\;\;\;\;\;\;\;\;\;NA\\
\hline
    II: Zeeman field-dominated&\hspace{0.45cm}\;\;\;$l_{\Omega}\ll l_{\tau}\ll
    l_{\alpha}$&\hspace{0.9cm}\;oscillation
    decay;&\hspace{0.1cm}\;\;\;\;\;single exponential
    decay;\;\;\;\;\;\\
    moderate scattering
    regime&&\hspace{0.9cm}\;\;\;\;\;\;\;\;\;\;$\sqrt{2}l_{\tau}$&\hspace{0.85cm}\;\;\;\;\;\;$l_{\tau}l_{\alpha}/(\sqrt{3}l_{\Omega})$\\
\hline
    III: SOC-dominated moderate&\hspace{0.45cm}\;\;\;$l_{\alpha}\ll l_{\tau}\ll
    l_{\Omega}$
&\hspace{0.1cm}\;\;\;\;\;single exponential
decay;\;\;\;\;\;&\hspace{0.1cm}\;\;\;\;\;single exponential decay;\;\;\;\;\;\\
    scattering
    regime&&\hspace{0.9cm}\;\;\;\;\;$\sqrt{3}l_{\tau}l_{\Omega}/(2l_{\alpha})$
    &\hspace{0.85cm}\;\;\;\;\;\;\;\;\;\;\;\;$l_{\alpha}$\\
\hline
    IV: relatively strong scattering&\hspace{0.3cm}\;\;\;$l_{\tau}\ll l_{\alpha}, l_{\Omega} \&l_{\tau}\gg
    l_{\tau,y}^c$\;\;\;&\hspace{0.15cm}\;\;\;\;\;single exponential
    decay;\;\;\;\;\;&\hspace{0.1cm}\;\;\;\;\;single exponential 
    decay;\;\;\;\;\;\\
    regime&\;\;\;\;\;\;\;($l_{\tau}\ll l_{\alpha}\ll
    l_{\Omega}$)&\hspace{0.75cm}\;\;\;\;\;\;$\sqrt{3}l_{\tau}l_{\Omega}/(2l_{\alpha})$&
\hspace{0.85cm}\;\;\;\;\;\;\;\;\;\;\;\;$l_{\alpha}$\\
\hline
    V: strong scattering regime&\hspace{0.3cm}\;\;\;$l_{\tau}\ll l_{\alpha}, l_{\Omega} \&l_{\tau}\ll
    l_{\tau,y}^c$\;\;\;&\hspace{0.9cm}\;oscillation
    decay;&\hspace{0.1cm}\;\;\;\;\;single exponential 
    decay;\;\;\;\;\;\\
    &&\hspace{0.9cm}\;\;\;\;\;$\sqrt{\sqrt{3}l_{\tau}l_{\Omega}}$&
\hspace{0.85cm}\;\;\;\;\;\;\;\;\;\;\;\;$l_{\alpha}$\\
    \hline
    \hline
\end{tabular}\\
\hspace{-14.4cm}$l_{\tau,y}^c\approx 4l_{\alpha}^2/(\sqrt{3}l_{\Omega})$. 
\end{table}
\end{center}
\end{widetext}
 From Table~\ref{table_y}, it can be
  seen that the
  spin diffusion along the $\hat{\bf y}$-direction is divided into similar five
  regimes as the spin diffusion along the $\hat{\bf x}$-direction.  The anomalous
  behaviors for the spin diffusion along the $\hat{\bf x}$-direction also exist
  here. Nevertheless, new features arise in the spin diffusion along the $\hat{\bf
    y}$-direction. It is shown in Table~\ref{table_y} that in Regimes III, IV
  and V, the longitudinal spin diffusion length is only determined by the SOC
  oscillation length, but irrelevant to the scattering. This {\it robustness} to the
  scattering for the spin diffusion in a wide range is further revealed in the scattering
  dependence of the spin diffusion  
  (Sec.~\ref{scattering_dependence}).
 Below, it is found that based on the modified drift-diffusion model
 and modified inhomogeneous broadening picture, apart from Regime
IV, all the features in
different regimes can be obtained.

We first analyze Regime II, the Zeeman field-dominated moderate scattering
  regime ($l_{\Omega}\ll l_{\tau}\ll
    l_{\alpha}$) and Regime V, the strong scattering regime
 ($l_{\tau}\ll l_{\alpha}, l_{\Omega}\ll l_{\tau,y}^c$). In these two
  regimes, with strong Zeeman and weak spin-orbit coupled fields, the effective
  inhomogeneous broadening field is written as
\begin{eqnarray} 
\nonumber
{\bgreek \omega}^y_{\rm eff}({\bf k})&=&(m/k_y)\sqrt{\Omega^2 +\alpha^2
  k_x^2}\hat{\bf x}'\\
&\approx&
\big[m\Omega/k_y+m\alpha^2
  k_x^2/(2\Omega k_y)\big]\hat{\bf x}',
\label{effective_Omega_y}
\end{eqnarray}
with $\hat{\bf x}'$ nearly parallel to $\hat{\bf x}$.
 In Regime II, i.e., the Zeeman field-dominated moderate scattering regime
 ($l_{\Omega}\ll l_{\tau}\ll
    l_{\alpha}$), when the
spin polarization is perpendicular to the Zeeman field (transverse
configuration), in the spacial domain, 
the spin vectors precess approximately around $(m\Omega/k_y)
\hat{\bf x}'$, which causes spacial oscillations whose
 period is proportional to $\langle |k_y|\rangle/(m\Omega)$ [$l_T^y\approx \sqrt{3/2}l_{\Omega}$ in
Eq.~(\ref{oscillation_Sz_period})]. Moreover, due to the fast spacial
oscillations, the spin memory is lost during one scattering, and hence the spin
diffusion length is the mean free path approximately ($L_T^y\approx
\sqrt{2}l_{\tau}$ in Table~\ref{table_y}). When the spin polarization is parallel to the Zeeman field (longitudinal
  configuration), the effective inhomogeneous broadening field ${\bgreek \omega}^y_{\rm eff}({\bf
    k})$ cannot cause spin precession efficiently and the steady-state spin polarization
  decays without any oscillation. From the modified drift-diffusion
  model,\cite{Peng43,Peng44,Peng45,Peng46,Peng47,Peng} as calculated in Sec.~\ref{analytical_x},
  $l_s\approx \sqrt{2}l_{\tau}l_{\alpha}/(\sqrt{3}l_{\Omega})$
 [$L_{L}^y\approx l_{\tau}l_{\alpha}/(\sqrt{3}l_{\Omega})$ in
 Table~\ref{table_y}].

In Regime V, i.e., the strong scattering regime ($l_{\tau}\ll
  l_{\alpha}, l_{\Omega}, l_{\tau,y}^c$), we analyze the limit situation with
  strong Zeeman and weak spin-orbit coupled fields ($l_{\tau}\ll
  l_{\alpha}, l_{\Omega}\ll l_{\tau,y}^c$). For the transverse
spin diffusion, the inhomogeneous
broadening is dominantly determined by the Zeeman field
[Eq.~(\ref{effective_Omega_y})].
 Hence, the transverse spin diffusion length is suppressed
by the Zeeman field, but less influenced by the SOC. Furthermore, during the
diffusion, the spin-conserving scattering suppresses the spin diffusion
 ($L_{L}^y\approx \sqrt{\sqrt{3}l_{\tau}l_{\Omega}}$ in
Table~\ref{table_y}). For the longitudinal spin diffusion, the inhomogeneous
broadening cannot cause the spin precession efficiently (single exponential
decay). According to the modified drift-diffusion model, the spin
diffusion length depends only on the SOC oscillation
length ($L_T^y\approx l_{\alpha}$ in
Table~\ref{table_x}).

We then analyze Regime III, SOC-dominated moderate scattering regime ($l_{\alpha}\ll
 l_{\tau}\ll l_{\Omega}$), and Regime IV, relatively strong scattering regime ($l_{\tau,y}^c\ll l_{\tau}\ll
  l_{\alpha}, l_{\Omega}$). With weak Zeeman and strong spin-orbit coupled
  fields, the effective inhomogeneous broadening field reads
\begin{eqnarray} 
\nonumber
{\bgreek \omega'}^y_{\rm eff}({\bf k})&=&(m/k_y)\sqrt{\alpha^2
  k_x^2+\Omega^2}\hat{\bf z}'\\
&\approx&
\big[m\alpha k_x/k_y+m\Omega^2/(2\alpha k_xk_y)\big]\hat{\bf z}',
\label{effective_alpha_y}
\end{eqnarray}
where $\hat{\bf z}'$ is parallel to $\hat{\bf z}$ approximately. In the
SOC-dominated moderate scattering regime (Regime III with $l_{\alpha}\ll
 l_{\tau}\ll l_{\Omega}$), in the
 transverse configuration, the spin polarization is nearly parallel to
 ${\bgreek \omega'}^y_{\rm eff}({\bf k})$. Therefore, the effective
 inhomogeneous broadening cannot cause the spin precession effectively (single
 exponential decay).
  From the modified drift-diffusion model,
 the diffusion length is proportional to
 $l_{\Omega}l_{\tau}/l_{\alpha}$, which is consistent with $L_y^T\approx
 \sqrt{3}l_{\Omega}l_{\tau}/(2l_{\alpha})$ in Table~\ref{table_y}.
 In the longitudinal configuration, the steady-state spin polarization
is perpendicular to ${\bgreek \omega'}^y_{\rm eff}({\bf k})$ approximately.
 One notes that in ${\bgreek \omega'}^y_{\rm eff}({\bf k})$, the spin-orbit coupled
field $(m\alpha k_x/k_y)\hat{\bf z}'$ provides the
dominant inhomogeneous broadening. Moreover, this inhomogeneous broadening not
only depends on $k_y$ but also $k_x$, which can be more efficient than the one
 depending only on $k_x$ or $k_y$. Due to this efficient inhomogeneous
 broadening,
 the steady-state spin polarization decays due to the interference
without oscillation. Moreover, the
spin memory can be lost in the scale of SOC oscillation length, which cannot persist
in the mean free path ($L_{L}^y\approx l_{\alpha}$ in Table~\ref{table_y}).
 This is different from the transverse spin diffusion
in the Zeeman filed-dominated moderate scattering regime ($l_{\Omega}\ll
 l_{\tau}\ll l_{\alpha}$).

In the relatively strong scattering regime
 (Regime IV with $l_{\tau,y}^c\ll l_{\tau}\ll
  l_{\alpha}, l_{\Omega}$). From Table~\ref{table_y},
 one observes that the behaviors for both the transverse and longitudinal spin
diffusions are similar to the ones in Regime III, i.e., the SOC-dominated moderate scattering
regime ($l_{\alpha}\ll l_{\tau}\ll l_{\Omega}$). We address that this
  behavior is hard to be understood from both the modified drift-diffusion model
and modified inhomogeneous broadening picture. For the transverse spin diffusion, the spin polarization is
 nearly parallel to $(m\alpha k_x/k_y)\hat{\bf z}'$, and hence the spin diffusion length
 is $\sqrt{2}l_{\alpha}/\sqrt{3}$ from the modified drift-diffusion
 model. For the longitudinal spin
 diffusion, the spin polarization is perpendicular to the inhomogeneous
 broadening field $(m\alpha k_x/k_y)\hat{\bf z}'$ approximately, and hence the
 SOC can suppress the
 spin diffusion.
 Moreover, the
 spin-conserving scattering can suppress the longitudinal spin diffusion. From
 this analysis, the longitudinal spin diffusion length is suppressed by the SOC
 strength and scattering, but irrelevant to the Zeeman field (modified
 inhomogeneous broadening picture).
 However, the
pictures above fail to explain the transverse and longitudinal spin diffusion
lengths for Regime IV in
Table~\ref{table_y} with $L_{T}^y\approx
\sqrt{3}l_{\tau}l_{\Omega}/(2l_{\alpha})$ and $L_{L}^y\approx l_{\alpha}$.
 This is because for the transverse situation, although the modified drift
   diffusion model can explain the spin diffusion along the ${\hat{\bf
       x}}$-direction, it is too rough to 
 consider the anisotropy between the diffusions along the $\hat{\bf x}$- and
 $\hat{\bf y}$-directions in the relatively strong scattering regime.
 This was first pointed out by Zhang and Wu in the
   study of the spin diffusion in graphene.\cite{Peng} For the longitudinal spin diffusion, when the scattering is
strong, the shortest length scale in the spin diffusion is the mean free
path. In this situation, there exists strong competition
 between the effective inhomogeneous broadening and scattering, which makes the
 behavior of the spin diffusion complicated.\cite{Peng,JinLuo_infinite,double}

Finally, we emphasize that from Table~\ref{table_y}, for the transverse spin
  diffusion along the $\hat{\bf y}$-direction, in Regimes III, IV and V,
  the spin diffusion lengths are irrelevant to the scattering. Therefore, with
  weak Zeeman and strong spin-orbit coupled fields ($l_{\alpha}\ll l_{\Omega}$),
a specific situation can be realized where the spin diffusion length is robust
against the scattering except with the extremely weak scattering.
 It is noted that in the strong scattering regime,  
    this feature was
    predicted in the simple drift-diffusion
    model\cite{Peng43,Peng44,Peng45,Peng46,Peng47} and was also revealed
    in graphene by the KSBE approach.\cite{Peng} In this work, we have further extended
    it into the weak scattering
  regime.

\section{Numerical results}
\label{numerical}
In the numerical calculation,
 the KSBEs are solved by employing the double-side boundary
conditions,\cite{double}
\begin{equation}
\left\{\begin{array}{cc}\rho_{{\bf k}}(\zeta=0,t)=\frac{\displaystyle f_{{\bf k}\uparrow}+f_{{\bf
      k}\downarrow}}{\displaystyle 2}+\frac{\displaystyle f_{{\bf k}\uparrow}-f_{{\bf
      k}\downarrow}}{\displaystyle 2}{\bgreek \sigma}\cdot \hat{\bf n},\hspace{0.15cm} k_{\zeta}>0\\
\hspace{-0.03cm}\rho_{{\bf k}}(\zeta=L,t)=f^0_{\bf k},\hspace{4cm}k_{\zeta}<0
\end{array}\right.,
\label{initial}
\end{equation}  
where $f_{{\bf k}\sigma}=\{\exp[(\varepsilon_{{\bf
    k}}-\mu_{\sigma})/(k_BT)]+1\}^{-1}$
 is the Fermi distribution function at temperature $T$,
 with $\mu_{\uparrow,\downarrow}$ standing for the chemical potentials determined by the atom
density $n_a$=$\sum_{\bf k}$Tr[${\rho_{\bf k}}$] and the 
spin polarization $P(0)$ in the left part of the system; $\hat{\bf n}$ denotes the spin
  polarization direction; $f^0_{\bf k}$ is the Fermi distribution at equilibrium.
When the system evolves to the steady state, one
obtains the diffusion length from the spatial evolution of the spin polarization
$P(\zeta)=\sum_{\bf k}\mbox{Tr}[\rho_{\bf
    k}(\zeta) {\bgreek \sigma}\cdot \hat{\bf n}]/n_a$.

Within the experimental
feasibility by referring to
the experiment by Wang {\em et al.},\cite{Fermi_SOC_1} 
 the parameters are chosen as follows. The lowest two magnetic sublevels $|9/2,9/2\rangle$ and
$|9/2,7/2\rangle$ are coupled by a pair of Raman
beams with wavelength $\lambda=773$ nm and the frequency difference
$\omega/(2\pi)=10.27$ MHz. The Raman
detuning $\delta=\omega_z-\omega$ is set to be zero by choosing the Zeeman
shift $\omega_z/(2\pi)=10.27$ MHz.
 The recoil momentum and energy are set to be
$k_r=k_0/10$ with $k_0=2\pi/\lambda$, and hence $E_r=k_r^2/(2m)=2\pi\times 83.4$
Hz. In our study, the SOC
strength varies from $0.5\alpha_0$ to $6\alpha_0$ with $\alpha_0=-2k_r/m$.
 The strengths of the Zeeman field $\Omega$ vary from 
 $10 E_r$ to $450E_r$.  Furthermore, the Fermi momentum is set
  to be $k_{\rm F}=30 k_r$.\cite{Fermi_SOC_1}
 It is noted that with these
parameters,
 the condition that the Zeeman and SOC energies are much
smaller than the Fermi energy is satisfied.

 Moreover, the temperature is set
  to $T=0.3T_{\rm F}$ with $T_{\rm F}$ being the
  corresponding Fermi temperature.\cite{Fermi_SOC_1} With these parameters,
  the thermal deBroglie wavelength $\Lambda_{\rm dB}=h/\sqrt{2\pi mk_BT}\approx
  0.26~\mu{\rm m}$.  
  For the 3D isotropic speckle disorder, $V_R/k_B=1250$~nK and
 $\sigma_R=0.27~\mu{\rm m}$.\cite{Disorder_3D_Fermion,Disorder_3D_Boson} With
 these disorder parameters, the mean free path $l_{\tau}\approx 5~{\mu}{\rm m}$. In
 our study, the strength of the disorder strength $V$ is tuned by the
 laser.\cite{Disorder_3D_Fermion,Disorder_3D_Boson,Disorder_1D_Boson,Disorder_theory,Configuration}
 One notes that when $(V/V_R)^2\gtrsim 20$, $l_{\tau}\lesssim
 \Lambda_{\rm dB}$. According to the Ioffe-Regel condition for the Anderson
 localization,\cite{AL_condition} the Anderson
 localization may become relevant. Nevertheless,
 in our study, to compare with the analytical
   results in different
regimes revealed in Sec.~\ref{analytical}, the numerical 
calculations are extended to $(V/V_R)^2\gg 20$.

\subsection{Scattering strength dependence}
\label{scattering_dependence}
\begin{widetext}
\begin{figure}[htb]
  \begin{minipage}[]{18cm}
    \hspace{0 cm}\parbox[t]{7.5cm}{
      \includegraphics[width=7.45cm,height=12.1cm]{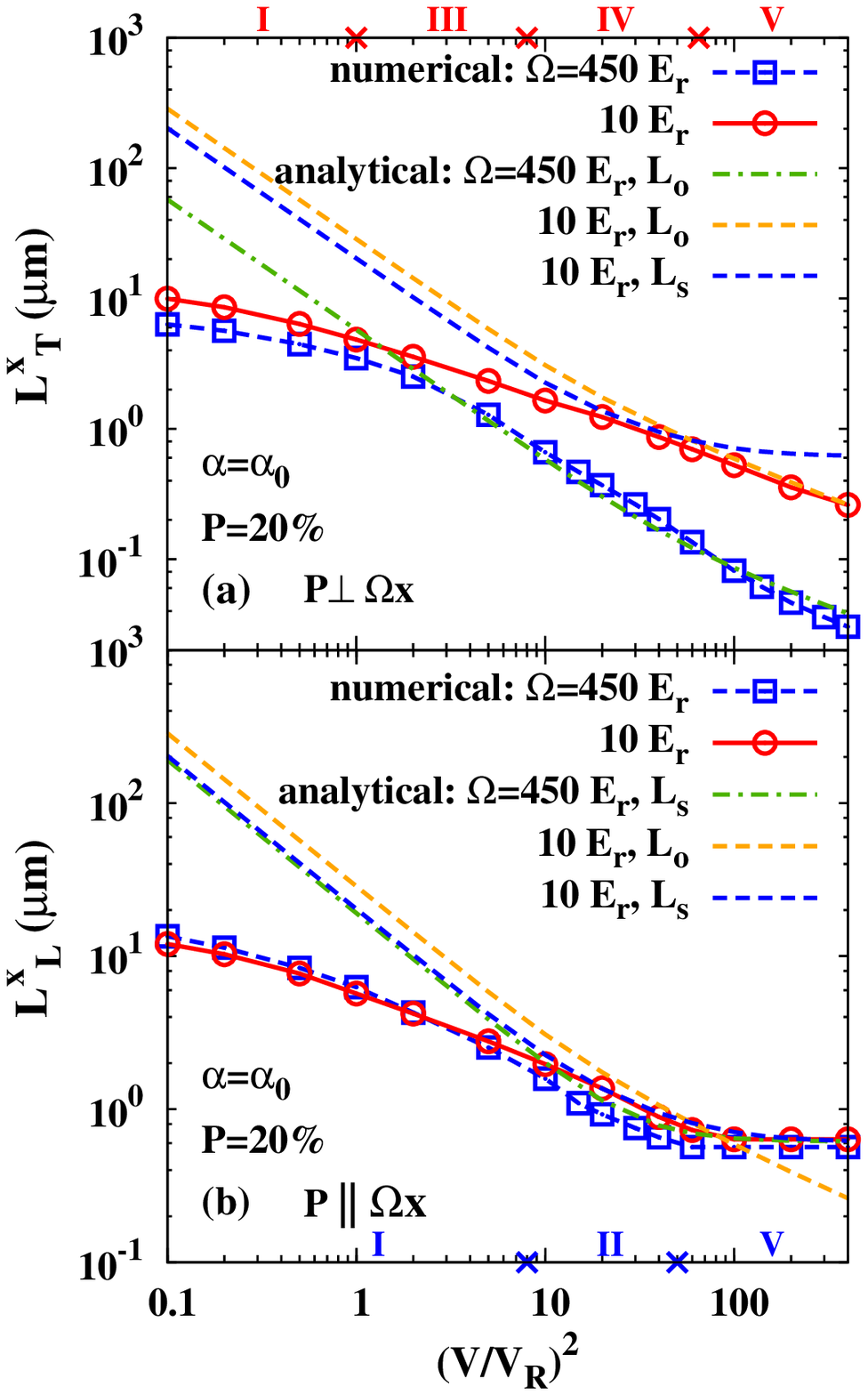}}
    \hspace{0cm}\parbox[t]{7.5cm}{
      \includegraphics[width=7.45cm,height=12.1cm]{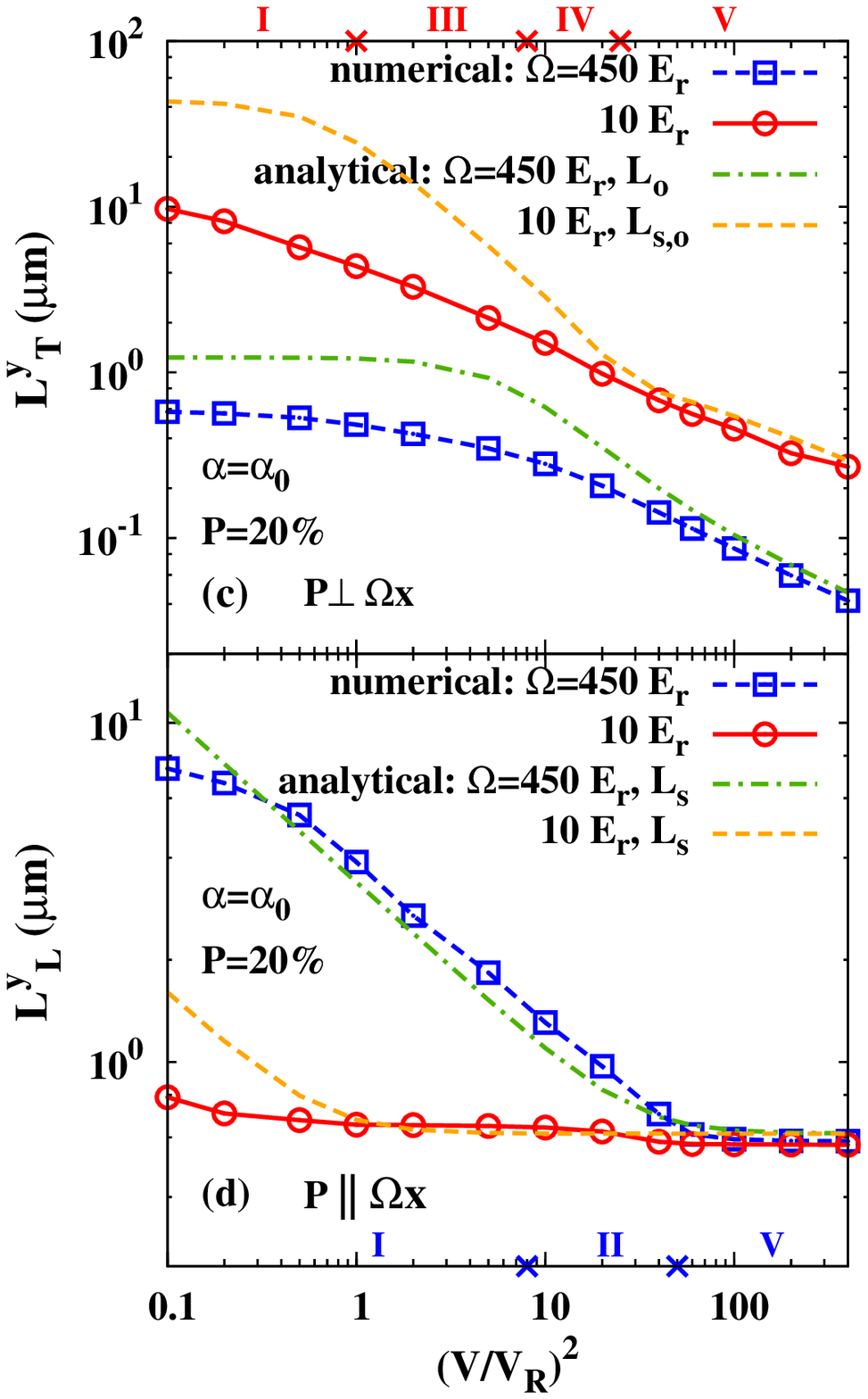}}
  \end{minipage}
\begin{minipage}[]{18cm}
\begin{center}
  \caption{(Color online) Scattering strength dependence of the steady-state
spin diffusion of spin-orbit coupled $^{40}$K gas with the 3D isotropic speckle
disorder. The SOC strength $\alpha=\alpha_0$ and the spin polarization $P=20\%$. For the spin diffusion along the
 $\hat{\bf x}$-direction ($\hat{\bf
  y}$-direction), the transverse and longitudinal spin diffusion
lengths $L_{T}^x$ and $L_{L}^x$ ($L_{T}^y$ and $L_{L}^y$) are
shown in Figs.~\ref{figyw3}(a) and (b) [Figs.~\ref{figyw3}(c) and (d)], respectively. Situations
with strong ($\Omega=450 E_r$) and weak ($\Omega=10 E_r$) Zeeman fields are
calculated both analytically and numerically. The blue (red) crosses on the frames indicate
 the boundaries between different regimes represented by the blue (red) Roman
  numbers at the lower (upper) frame when
 $\Omega=450 E_r$ ($10 E_r$). It is shown that the analytical
calculations agree with the numerical ones in the relatively strong/strong 
scattering regime and crossover region between the relatively strong and moderate
  scattering regimes.}
\label{figyw3}
\end{center}
\end{minipage}
\end{figure}
\end{widetext}

In this part, we study the scattering strength dependence of the steady-state
spin diffusion of spin-orbit coupled $^{40}$K gas in the 3D isotropic speckle
disorder. The SOC strength is set to be $\alpha_0$ and the spin
  polarization is chosen to be $P=20\%$.
 For the spin diffusion along the $\hat{\bf x}$-direction (${\hat {\bf
    y}}$-direction),
both the transverse and longitudinal spin diffusion
lengths $L_{T}^x$ and $L_{L}^x$ ($L_{T}^y$ and $L_{L}^y$) are
shown in Figs.~\ref{figyw3}(a) and (b) [Figs.~\ref{figyw3}(c) and (d)].

We first analyze the spin diffusion along the $\hat{\bf x}$-direction
  [Figs.~\ref{figyw3}(a) and (b)].
  For the transverse
  spin diffusion, it can be seen from
  Fig.~\ref{figyw3}(a) that no matter the Zeeman
  field is strong with $\Omega=450 E_r$ (the blue dashed curve with squares) or
  weak with $\Omega=10 E_r$ (the red solid curve with circles), the transverse spin
  diffusion length $L_T^x$ decreases with the increase of the disorder
  strength. The
  underlying physics can be understood as follows. We first calculate the
  characteristic lengths for the system defined in Sec.~\ref{analytical}: the SOC
  oscillation length $l_{\alpha}\approx
0.6$~${\mu}{\rm m}$; the Zeeman
  oscillation length $l_{\Omega}\approx 0.1$~${\mu}{\rm m}$ (4~${\mu}{\rm m}$)
 for $\Omega=450 E_r$ ($10 E_r$); when $\Omega=450 E_r$ ($10 E_r$), the
   crossover length between the relatively strong and strong scattering regimes
 $l_{\tau,x}^c\approx\sqrt{3}l_{\alpha}^2/(2l_{\Omega})\approx 3$~${\mu}{\rm m}$
 (0.08~${\mu}{\rm m}$), i.e., $(V/V_R)^2_c\approx 2$ [$(V/V_R)^2_c\approx 65$].
 Furthermore, it
is calculated that when $l_{\tau}\approx l_{\alpha}$, $(V/V_R)^2\approx 8$; when
$l_{\tau}\approx l_{\Omega}$, $(V/V_R)^2\approx 50$ [$(V/V_R)^2\approx1$] for $\Omega=450 E_r$
($10 E_r$). Accordingly, with the
increase of the disorder strength, the system experiences several regimes. For
$\Omega=450E_r$, the regimes are approximately divided into
\begin{equation}
\left\{\begin{array}{cc}
\hspace{-0.88cm}{\rm I}:~l_{\tau}\gtrsim l_{\Omega}, l_{\alpha},\hspace{0.75cm}{\rm
  when}\hspace{0.1cm}
 (V/V_R)^2\lesssim 8\\
\hspace{-0.1cm} {\rm II}: l_{\Omega}\lesssim
l_{\tau}\lesssim l_{\alpha},\hspace{0.47cm}{\rm
  when}\hspace{0.1cm}8 \lesssim (V/V_R)^2\lesssim 50\\
\hspace{-0.65cm} {\rm V}: l_{\tau}\ll l_{\Omega}, l_{\alpha}, l_{\tau,x}^c\hspace{0.2cm}{\rm
  when}\hspace{0.1cm}(V/V_R)^2\gg 50
\end{array}\right.,
\label{regime_450}
\end{equation}
which are labelled by the blue Roman numbers with the boundaries
  indicated by the blue crosses at the lower frame of Fig.~\ref{figyw3}(b).
Therefore, when $(V/V_R)^2\lesssim 8$, the system is in Regime I,
 and our calculation shows that the transverse diffusion length decreases with the
increase of the disorder strength.
When $8
\lesssim (V/V_R)^2\lesssim 50$ [$(V/V_R)^2\gg 50$], the system lies in Regime
II (V), and from Table~\ref{table_x}, one
comes to $L^x_{T}\approx
2l_{\tau}/\sqrt{3}$ ($L^x_{T}\approx\sqrt{2l_{\tau}l_{\Omega}/\sqrt{3}}$)
with the steady-state spin polarization showing oscillation decay.
Hence, 
the transverse spin diffusion length decreases with the increase of the disorder
strength.
One notes that the corresponding results calculated from the analytical formula Eq.~(\ref{Lox}) (the
green dot-dashed curve) agree with the numerical ones in Regimes II and V in
Fig.~\ref{figyw3}(a).  
For $\Omega=10E_r$, the regimes are approximately divided into
\begin{equation}
\hspace{-0.1cm}\left\{\begin{array}{cc}
\hspace{-0.88cm}{\rm I}:~~l_{\tau}\gtrsim l_{\Omega}, l_{\alpha},\hspace{1.05cm}{\rm
  when}\hspace{0.1cm}
 (V/V_R)^2\lesssim 1\\
\hspace{-0.26cm}{\rm III}: l_{\alpha}\lesssim
l_{\tau}\lesssim l_{\Omega},\hspace{0.7cm}{\rm
  when}\hspace{0.1cm}1 \lesssim (V/V_R)^2\lesssim 8\\
\hspace{-0.03cm} {\rm IV}: l_{\tau,x}^c\lesssim l_{\tau}\lesssim l_{\Omega}, l_{\alpha},\hspace{0.05cm}{\rm
  when}\hspace{0.1cm}8\lesssim (V/V_R)^2\lesssim 65\\
\hspace{-0.51cm} {\rm V}:~l_{\tau}\ll l_{\Omega}, l_{\alpha}, l_{\tau,x}^c,\hspace{0.3cm}{\rm
  when}\hspace{0.1cm}(V/V_R)^2\gg 65
\end{array}\right.,
\label{regime_10}
\end{equation}
which are shown by the red Roman numbers with the boundaries indicated by
  the red crosses at the upper frame of Fig.~\ref{figyw3}(a).
Specifically, in Regime I when
$(V/V_R)^2\lesssim 1$, it is shown that the transverse spin diffusion length decreases with the increase of the
scattering strength. In Regime III (V)
 with $1 \lesssim (V/V_R)^2\lesssim 8$ [$(V/V_R)^2\gg 65$], 
from Table~\ref{table_x}, 
 it is obtained that $L^x_{T}\approx l_{\tau}l_{\Omega}/(\sqrt{3}l_{\alpha})$
 ($L^x_{T}\approx\sqrt{2l_{\tau}l_{\Omega}/\sqrt{3}}$) with the steady-state
 spin polarization being single exponential (oscillation) decay. Hence, the transverse 
 spin diffusion length decreases with the increase of disorder strength.
 One further notices that in Fig.~\ref{figyw3}(a),
 the result calculated from the analytical formula Eq.~(\ref{Lox})
 (the orange dashed curve) agrees with the numerical
 one in Regime V.

For the longitudinal spin diffusion
  along the $\hat{\bf x}$-direction, it can be seen from
Fig.~\ref{figyw3}(b) that no matter the Zeeman field is strong ($450 E_r$, blue dashed curve with
 squares)
or weak ($10 E_r$, red solid curve with circles), with the increase of the disorder strength,
 the spin diffusion length decreases first and then becomes
insensitive to the scattering. This can be understood as
follows. With the increase of the disorder strength, the corresponding regimes
can also be 
divided according to Eq.~(\ref{regime_450}) [Eq.~(\ref{regime_10})]
 when $\Omega=450 E_r$ ($10 E_r$). Specifically, when $(V/V_R)^2\lesssim 8$
 [$(V/V_R)^2\lesssim 1$]
 for $\Omega=450 E_r$ ($10
 E_r$), the system is in  Regime I,  with the longitudinal spin diffusion
 suppressed by the scattering. When $8
\lesssim (V/V_R)^2\lesssim 50$ [$1 \lesssim (V/V_R)^2\lesssim 8$] for
$\Omega=450E_r$ ($10E_r$),
 the system lies in Regime II (III), and one comes to $L^x_{L}\approx
l_{\tau}l_{\alpha}/(\sqrt{3}l_{\Omega})$
 [$L^x_{L}\approx \sqrt{2}l_{\tau}l_{\Omega}/(\sqrt{3}l_{\alpha})$] with the
 steady-state spin polarization showing single exponential (oscillation) decay. When
$(V/V_R)^2\gg 50$ [$(V/V_R)^2\gg 65$] for $\Omega=450 E_r$ ($10
 E_r$), the system lies in Regime V, it is obtained that $L^x_{L}\approx
l_{\alpha}$ (single exponential decay). Therefore,  
the longitudinal spin diffusion is suppressed first and then becomes insensitive
to the scattering
with the
increase of the disorder strength. Also in
Fig.~\ref{figyw3}(b), for
$\Omega=450E_r$, it is shown that the results 
 calculated from the analytical formula Eq.~(\ref{Lsx}) (the green dot-dashed curve) agrees with the
numerical one in both Regimes II and V; for
  $\Omega=10E_r$, the analytical results 
 (the blue dashed curve) agree with the
numerical ones in Regime V.

We then turn to the spin diffusion along the $\hat{\bf y}$-direction
  [Figs.~\ref{figyw3}(c) and (d)]. For the transverse spin diffusion, it is shown in 
  Fig.~\ref{figyw3}(c) that for both the strong ($450 E_r$, blue dashed curve with
 squares) and weak ($10 E_r$, red solid curve with circles) Zeeman fields,
 the spin diffusion length decreases with the increase of the 
  disorder strength.  It is calculated that for $\Omega=450 E_r$
($10 E_r$),  
$l_{\tau,y}^c\approx 8.3~{\mu{\rm m}}$ ($0.2~{\mu{\rm m}}$),
 i.e., $(V/V_R)^2_c\approx 0.6$ [$(V/V_R)^2_c\approx 25$]. Accordingly, 
  when the Zeeman
  field is strong ($\Omega=450 E_r$), one can divide the regimes according to
  Eq.~(\ref{regime_450}).
Specifically, when 
$(V/V_R)^2\lesssim 8$,
the transverse spin diffusion is calculated to be suppressed by the
scattering. When $8\lesssim (V/V_R)^2\lesssim 50$
[$(V/V_R)^2\gg 50$], one obtains from
 Table~\ref{table_y} that $L_T^{y}\approx
  \sqrt{2}l_{\tau}$ ($L_T^{y}\approx
  \sqrt{\sqrt{3}l_{\Omega}l_{\tau}}$) with the steady-state spin polarization
  showing oscillation decay.
Therefore, with the increase of the scattering strength,
  the transverse spin diffusion is suppressed.
 When the Zeeman field is weak ($10 E_r$),
  the regimes are approximately divided into  
\begin{equation}
\hspace{-0.2cm}\left\{\begin{array}{cc}
\hspace{-0.82cm}{\rm I}:~~l_{\tau}\gtrsim l_{\Omega}, l_{\alpha},\hspace{1.05cm}{\rm
  when}\hspace{0.1cm}
 (V/V_R)^2\lesssim 1\\
\hspace{-0.2cm} {\rm III}: l_{\alpha}\lesssim
l_{\tau}\lesssim l_{\Omega},\hspace{0.75cm}{\rm
  when}\hspace{0.1cm}1 \lesssim (V/V_R)^2\lesssim 8\\
\hspace{0.02cm} {\rm IV}: l_{\tau,y}^c\lesssim l_{\tau}\lesssim l_{\Omega}, l_{\alpha},\hspace{0.1cm}{\rm
  when}\hspace{0.1cm}8\lesssim (V/V_R)^2\lesssim 25\\
\hspace{-0.45cm} {\rm V}:~l_{\tau}\ll l_{\Omega}, l_{\alpha}, l_{\tau,y}^c,\hspace{0.3cm}{\rm
  when}\hspace{0.1cm}(V/V_R)^2\gg 25
\end{array}\right.,
\label{regime_10_y}
\end{equation}
which are labelled by the red Roman numbers with the boundaries
  indicated by the red crosses at the upper frame of Fig.~\ref{figyw3}(c).
Specifically, when $(V/V_R)^2\lesssim 1$, the
transverse spin diffusion length decreases with the increase of the disorder
strength. When $1 \lesssim (V/V_R)^2\lesssim 8$
[$(V/V_R)^2\gg 25$], it is obtained from Table~\ref{table_y} that $L_T^{y}\approx
  \sqrt{3}l_{\Omega}l_{\tau}/(2l_{\alpha})$ ($L_T^{y}\approx
  \sqrt{\sqrt{3}l_{\Omega}l_{\tau}}$) with the steady-state spin polarization
  being single exponential (oscillation) decay. Therefore, the transverse spin diffusion length decreases
  with the increase of the disorder strength. Furthermore, it can be seen in
    Fig.~\ref{figyw3}(c) that the results
    calculated from the analytical formulas Eqs.~(\ref{Ltyminus})
 and (\ref{Lty}) agree with the numerical ones in Regime V.

For the longitudinal spin diffusion
  along the $\hat{\bf y}$-direction, it is shown in
  Fig.~\ref{figyw3}(d) that with the increase of the scattering strength, for both
  the strong ($450 E_r$) and weak ($10 E_r$) Zeeman fields, the
  longitudinal spin
  diffusion length is suppressed first and then become insensitive to the
  scattering. Specifically, when $\Omega=10 E_r$ ($l_{\alpha}\ll l_{\Omega}$),
 the longitudinal spin diffusion is robust against the scattering 
except with extremely weak scattering. Here, for the
  strong (weak) Zeeman field $\Omega=450 E_r$ ($10 E_r$), the regimes are
  divided according to Eq.~(\ref{regime_450}) [Eq.~(\ref{regime_10_y})].
 For the strong Zeeman field ($450 E_r$), when $(V/V_R)^2\lesssim 8$,
 the longitudinal spin diffusion is suppressed by the
 scattering. When
 $8<(V/V_R)^2\lesssim 50$ [$(V/V_R)^2\gg 50$], it is obtained from
 Table~\ref{table_y} that $L_{L}^y\approx
  l_{\tau}l_{\alpha}/(\sqrt{3}l_{\Omega})$ ($L_{L}^y\approx l_{\alpha}$) with
  the steady-sate spin polarization being single exponential decay.
 Hence, the longitudinal spin diffusion length decreases first and then become
insensitive to the scattering 
with the increase of the disorder strength. For the weak Zeeman field
($10 E_r$), when $(V/V_R)^2\lesssim 1$, our
calculation shows that the longitudinal spin diffusion length decreases slowly
with the increase of the disorder strength. When $(V/V_R)^2\gtrsim 1$,
 one obtains from Table~\ref{table_y} that 
$L_{L}^y\approx l_{\alpha}$ (single exponential decay).
 Hence, the longitudinal spin diffusion length is
 insensitive to the scattering in a wide range. Moreover, it is shown in
    Fig.~\ref{figyw3}(d) that the results
    calculated from the analytical formula Eq.~(\ref{Lly})
  agree with the numerical ones in both the moderate and strong scattering regimes.

Finally, we address the specific features in the scattering strength dependence of
  the transverse and longitudinal spin diffusions along the $\hat{\bf x}$- and
  $\hat{\bf y}$-directions. Our calculations show that in the weak
 ($l_{\tau}\gtrsim l_{\Omega}, 
l_{\alpha}$) 
  scattering
 regime (Regime I), both the transverse and
  longitudinal spin diffusions are 
  suppressed by the scattering. In the strong scattering limit ($l_{\tau}\ll l_{\Omega},
l_{\alpha},l_{\tau}^c$), the longitudinal spin
 diffusions along the $\hat{\bf x}$ and $\hat{\bf
   y}$-directions are insensitive to the scattering. Specifically, when $l_{\alpha}\ll l_{\Omega}$,
 the longitudinal spin diffusion along the $\hat{\bf y}$-direction is robust against the scattering 
except with extremely weak scattering. We emphasize that this robust spin
  diffusion cannot be obtained from the over-simplified drift-diffusion
  model, where $l_s=\sqrt{D\tau_s({\bf
      k})}$ with $D=v_F^2\tau_k/3$.\cite{Peng43,Peng44,Peng45,Peng46,Peng47,Peng}  From the
  drift-diffusion model, in Regime V with $\tau_s({\bf k})\propto \tau_k^{-1}$,
 one surely obtains that the spin diffusion length is
  irrelevant to the scattering. However, in Regimes I and II, i.e., the weak
  scattering regime defined in the conventional DP spin
  relaxation,\cite{DP,Anomalous_DP_1,wu-review} $\tau_s({\bf k})\approx \tau_k$. Hence it is obtained that $l_s\propto
 \tau_k$ with the spin diffusion length suppressed by the scattering. 
 One further notes that the experimental
  condition can be easily realized to observe this robust spin diffusion. On one
  hand, the spin diffusion is set to be along the $\hat{\bf y}$-direction with
  the initial spin
  polarization parallel to the Zeeman field; on the other hand, the
  Zeeman field is tuned to be much weaker than the spin-orbit coupled one by the
  laser field.

\subsection{Zeeman field dependence}
\label{Zeeman_dependence}
 
\begin{figure}[ht]
  {\includegraphics[width=7.5cm]{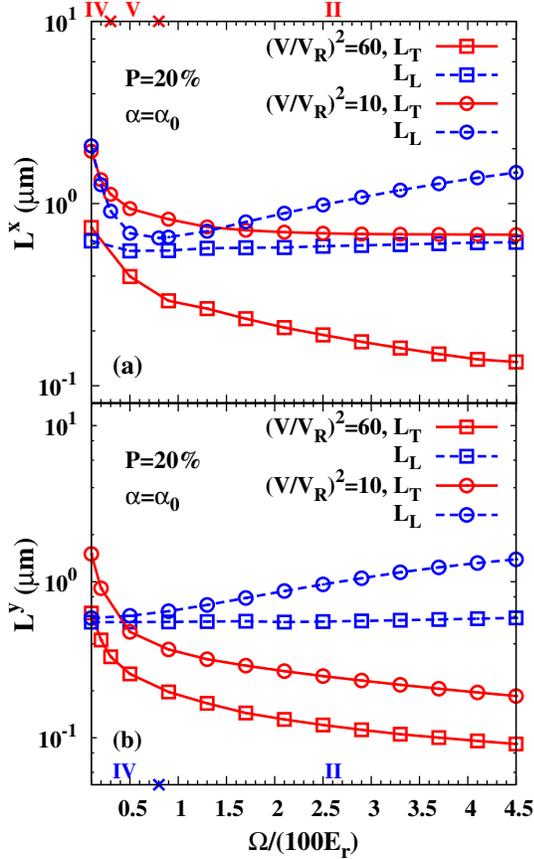}}
  \caption{(Color online) Zeeman field dependence of the transverse and
longitudinal spin diffusions along the (a) $\hat{\bf x}$- and (b) $\hat{\bf
  y}$-directions. The SOC strength $\alpha=\alpha_0$ and the spin
polarization $P=20\%$. Both the situations with the scattering strength
  $(V/V_R)^2=60$ (squares) and $(V/V_R)^2=10$ (circles) calculated
  numerically are presented. The green (blue) Roman numbers at the upper (lower)
  frame represent the
    different regimes when $(V/V_R)^2=10$ for the spin diffusion along the
      $\hat{\bf x}$-direction ($\hat{\bf y}$-direction) with the boundaries indicated by the
    green (blue) crosses.}
  \label{figyw4}
\end{figure}

In this part, we address the Zeeman field dependence of the transverse and
longitudinal spin diffusions along the $\hat{\bf x}$- [Fig.~\ref{figyw4}(a)]
and $\hat{\bf y}$-directions [Fig.~\ref{figyw4}(b)]. The SOC strength is set to be $\alpha_0$ and the spin
  polarization is chosen to be $P=20\%$. Here, we mainly
  address the specific features in the Zeeman field dependence when the system
  lies in the moderate and
strong scattering regimes, which can be realized by setting $(V/V_R)^2=60$ and
$(V/V_R)^2=10$, respectively.

We first focus on the case with $(V/V_R)^2=60$. When $(V/V_R)^2=60$, one
  observes from Figs.~\ref{figyw4}(a) and (b) that for
  the spin diffusion along both $\hat{\bf x}$- and $\hat{\bf y}$-directions,
 the transverse diffusion length decreases with the increase
of the Zeeman field, as shown by the red solid curve with
squares. This is because when $(V/V_R)^2=60$, for the spin diffusion along the 
$\hat{\bf x}$-direction ($\hat{\bf y}$-direction), $l_{\tau}\lesssim l_{\alpha},
l_{\Omega}, l_{\tau,x}^c$ ($l_{\tau}\lesssim l_{\alpha},
l_{\Omega}, l_{\tau,y}^c$)
 is satisfied. Hence, the system lies in Regime V. Accordingly, for the transverse
 spin diffusion along the $\hat{\bf x}$-direction ($\hat{\bf y}$-direction),
 one obtains from Table.~\ref{table_x} (Table.~\ref{table_y}) that 
 $L_{T}^x\approx \sqrt{2l_{\tau}l_{\Omega}/\sqrt{3}}$ [$L_{T}^y\approx
 \sqrt{3}l_{\Omega}l_{\tau}/(2l_{\alpha})$]. Therefore, with the increase of the Zeeman field,
the transverse spin diffusion length 
decreases. For the longitudinal spin
   diffusion, it is shown in Figs.~\ref{figyw4}(a) and (b) that no matter the spin
   diffusion is along the $\hat{\bf x}$-direction or the $\hat{\bf
     y}$-direction, the spin diffusion length is marginally influenced by the
   Zeeman field (blue dashed curve with squares). This anomalous behavior can be well understood from the
   analytical results in Regime V. For the longitudinal
 spin diffusion along the $\hat{\bf x}$-direction ($\hat{\bf y}$-direction), it is
 obtained that 
 $L_{L}^x\approx l_{\alpha}$ ($L_{L}^y\approx l_{\alpha}$). Accordingly, in Regime V, the
 longitudinal spin diffusions along both the $\hat{\bf x}$- and $\hat{\bf
   y}$-directions are
 marginally influenced by
 the Zeeman field. Based on the drift-diffusion model,\cite{Peng43,Peng44,Peng45,Peng46,Peng47,Peng}
 it is emphasized that this unique feature arises from the
insensitivity of diffusion coefficient and SRT to the Zeeman field in the strong
scattering limit.

We then analyze the case with $(V/V_R)^2=10$. We first divide the regimes
  for the spin diffusion along the $\hat{\bf x}$- and $\hat{\bf y}$-directions,
  respectively. For the spin diffusion along the $\hat{\bf x}$-direction, the
regimes for the system are divided into 
\begin{equation}
\left\{\begin{array}{cc}
\hspace{-1.25cm} {\rm IV}: l_{\tau,x}^c\lesssim l_{\tau}\lesssim l_{\Omega}, l_{\alpha},\hspace{0.35cm}{\rm
  when}\hspace{0.1cm}\Omega\lesssim 30 E_r\\
\hspace{0cm} {\rm V}:~l_{\tau}\lesssim l_{\Omega}, l_{\alpha}, l_{\tau,x}^c,\hspace{0.7cm}{\rm
  when}\hspace{0.1cm}30 E_r\lesssim\Omega\lesssim 80 E_r\\
\hspace{-1.2cm}{\rm II}:~l_{\Omega}\lesssim l_{\tau}\lesssim l_{\alpha},\hspace{1.07cm}{\rm
  when}\hspace{0.1cm}
 \Omega\gtrsim 80 E_r
\end{array}\right.,
\label{regime_Omega_2}
\end{equation}
which are labelled by the red Roman numbers with the boundaries indicated
by the red crosses at the upper frame of Fig.~\ref{figyw4}(a). For the spin diffusion along the 
$\hat{\bf y}$-direction, when $\Omega\lesssim 80 E_r$ and $\Omega\gtrsim 80 E_r$, the system lies in Regimes
IV and II, which are represented by the blue Roman numbers with the
boundaries indicated by the blue crosses at the lower frame of Fig.~\ref{figyw4}(b). It
is shown in Figs.~\ref{figyw4}(a) and (b) that in Regime II, the transverse
(longitudinal) diffusion length decreases slowly (increases) with the increase of the
Zeeman field, represented by the red solid (blue dashed) curve with circles.
 This is because for the transverse spin diffusion along the
 $\hat{\bf x}$-direction ($\hat{\bf y}$-direction), in Regime II, 
$L_{T}^x\approx 2l_{\tau}/\sqrt{3}$ ($L_{T}^y\approx
\sqrt{2}l_{\tau}$), with the diffusion length marginally influenced by the
Zeeman field. For the longitudinal spin diffusion, in Regime II, $L_{L}^x\approx
l_{\tau}l_{\alpha}/(\sqrt{3}l_{\Omega})$ [$L_{L}^y\approx
l_{\tau}l_{\alpha}/(\sqrt{3}l_{\Omega})$], leading to the enhancement of the
spin diffusion by the Zeeman field. We emphasize that this enhancement of the
spin diffusion arises from the suppression of the longitudinal spin relaxation
by the Zeeman field.

Finally, we emphasize the unique features in the Zeeman field dependence of the
  spin diffusions along the $\hat{\bf x}$- and $\hat{\bf y}$-directions, which can
  be observed in the experiment. These unique features arise in the longitudinal
situation, i.e., the initial spin polarization is parallel to the Zeeman field,
for the spin diffusion along both the $\hat{\bf x}$- and $\hat{\bf
  y}$-directions. On one hand, when the scattering is strong with the system in
Regime V, the longitudinal spin diffusion is marginally influenced by the Zeeman
field; on the other hand, when the scattering is relatively weak and the Zeeman
field is strong with the system in Regime II, the spin diffusion length is
enhanced by the Zeeman field. These unique features can be understood in the
framework of {\it modified} drift-diffusion model addressed in
 Sec.~\ref{analytical}.\cite{Peng43,Peng44,Peng45,Peng46,Peng47,Peng}
  It is emphasized that
 here the drift-diffusion model is applicable, which is very different from the
 longitudinal spin diffusion along the $\hat{\bf y}$-direction in Regimes III and
 IV [the red solid curve with circles in Fig.~\ref{figyw3}(d)] where the
 modified inhomogeneous broadening picture is used. We readdress
 that with the strong (weak) Zeeman and weak (strong) spin-orbit coupled fields, 
the condition to apply the modified drift-diffusion model/modified
  inhomogeneous broadening picure is that the initial spin
 polarization is parallel/perpendicular to the larger field.

\subsection{SOC strength dependence}
\label{SOC_dependence}
\begin{figure}[ht]
  {\includegraphics[width=7.6cm]{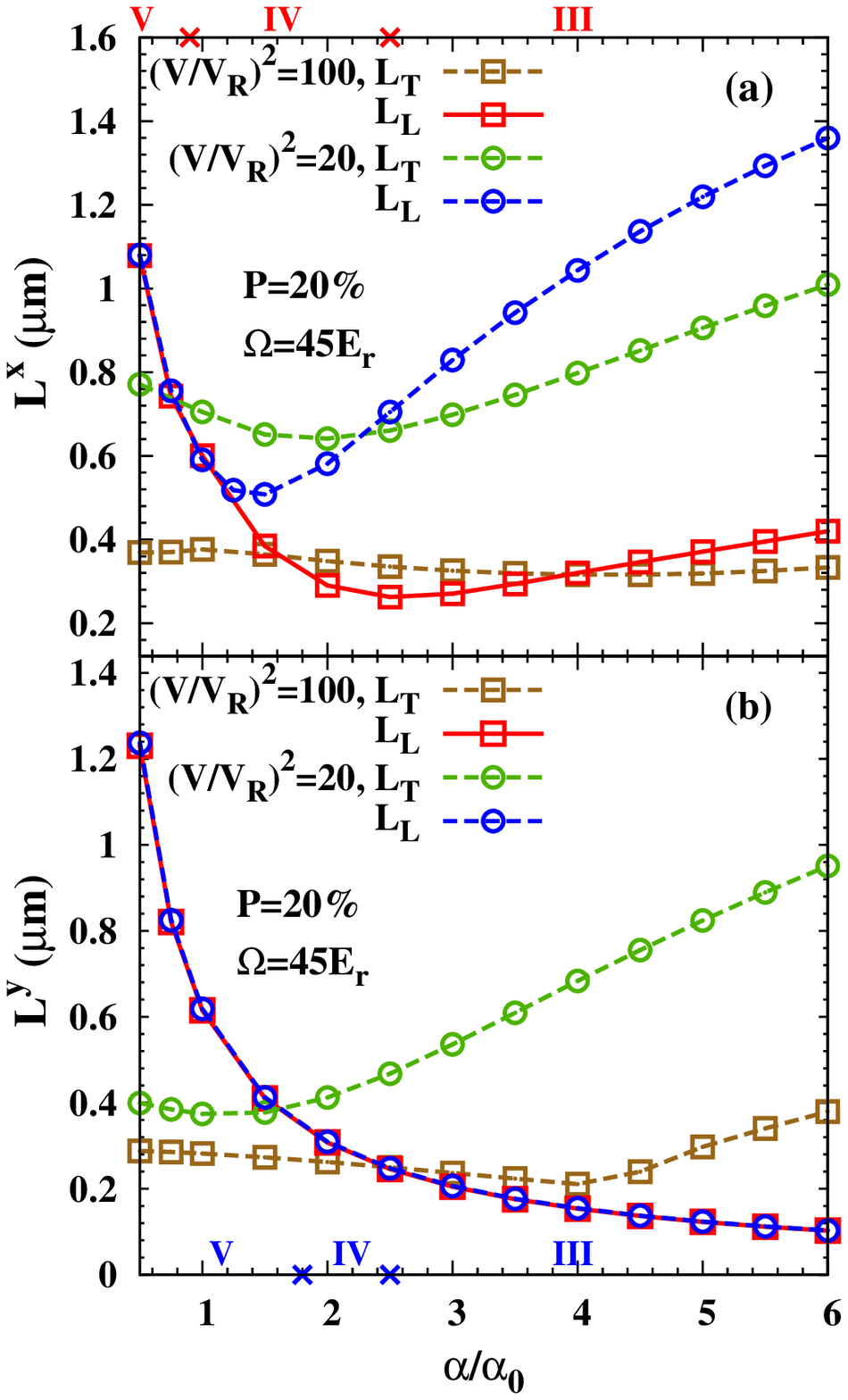}}
  \caption{(Color online) SOC dependence of the transverse and
longitudinal spin diffusions along the (a) $\hat{\bf x}$- and (b) $\hat{\bf
  y}$-directions. The Zeeman field $\Omega=45 E_r$ and the spin
polarization $P=20\%$. Both the situations with the scattering strength
  $(V/V_R)^2=100$ (squares) and $(V/V_R)^2=20$ (circles) calculated
  numerically are presented. The red (blue) Roman numbers at the upper (lower)
  frame represent the
    different regimes for the spin diffusion along the $\hat{\bf x}$-direction
 ($\hat{\bf y}$-direction) when $(V/V_R)^2=20$ with the boundaries indicated by the
    red (blue) crosses.}
  \label{figyw5}
\end{figure}

In this part, we analyze the SOC strength dependence of the transverse and
longitudinal spin diffusions along the $\hat{\bf x}$- [Fig.~\ref{figyw5}(a)]
and $\hat{\bf y}$-directions [Fig.~\ref{figyw5}(b)]. It has been well
  understood from the drift-diffusion
  model that in the strong
  scattering regime, the spin diffusion length is suppressed by
 the SOC ($l_{s}\propto
 1/|\alpha|$).\cite{Peng43,Peng44,Peng45,Peng46,Peng47,Peng} 
 In our study, besides the suppression
 of the spin diffusion length by the SOC, we find two unique
features in the SOC strength dependence which can not be derived from the over-simplified
drift-diffusion
  model: the spin diffusion length can be either
marginally influenced or even enhanced by the SOC. These features can be realized  when the system
lies in the moderate and strong scattering regimes. Accordingly, in our
calculation, the Zeeman field is set to
be $\Omega=45E_r$; the scattering strengths are set to be $(V/V_R)^2=100$ and $(V/V_R)^2=20$, respectively.

We first address the first unique feature, i.e., the marginal influence of the SOC on the spin
diffusion. In Figs.~\ref{figyw5}(a) and (b), one observes that when
$(V/V_R)^2=100$, no matter the spin diffusion is along the $\hat{\bf x}$-
or $\hat{\bf y}$-direction in the transverse configuration (the gray dashed
curve with squares), when $\alpha \lesssim 4\alpha_0$, the spin diffusion
length is marginally influenced by the SOC. This is because when $\alpha
\lesssim 4\alpha_0$, the system lies in Regime V with $L_T^x\approx
\sqrt{2l_{\tau}l_{\Omega}/\sqrt{3}}$ ($L_{L}^x\approx
\sqrt{\sqrt{3}l_{\tau}l_{\Omega}}$) for the transverse spin diffusion along the 
$\hat{\bf x}$-direction ($\hat{\bf y}$-direction). It is emphasized that
this robustness of the spin diffusion to the SOC cannot be obtained from the
drift-diffusion model.\cite{Peng43,Peng44,Peng45,Peng46,Peng47,Peng}
As we have addressed in Sec.~\ref{analytical}, in the transverse spin
 diffusion in Regime V, the inhomogeneous broadening has dominant influence on
 the spin diffusion.

We then analyze the second unique feature where the spin diffusion length
  can be enhanced by the SOC. For the spin diffusion along the $\hat{\bf
    x}$-direction, it is shown in Fig.~\ref{figyw5}(a) by the green (blue) dashed curve with
  circles that when $(V/V_R)^2=20$ with $\alpha\gtrsim
 2\alpha_0$, the transverse (longitudinal) spin diffusion is significantly
 enhanced by the SOC; for the spin diffusion along the $\hat{\bf
    y}$-direction, when $(V/V_R)^2=20$ with $\alpha\gtrsim
 2\alpha_0$ the transverse spin diffusion length also increases with the
  increase of the SOC (the green dashed curve with
  circles). This can be understood as follows. When $\alpha\gtrsim
 2\alpha_0$, the system lies in Regime III. Accordingly, for the transverse
 (longitudinal) spin diffusion along the $\hat{\bf x}$-direction, one obtains
 $L_T^x\approx l_{\tau}l_{\Omega}/(\sqrt{3}l_{\alpha})$ [$L_{L}^x\approx
 \sqrt{2}l_{\tau}l_{\Omega}/(\sqrt{3}l_{\alpha})$]; for the transverse spin
 diffusion along the $\hat{\bf y}$-direction, $L_T^x\approx
 \sqrt{3}l_{\tau}l_{\Omega}/(2l_{\alpha})$. This unique
 feature is in contrast to the prediction of the 
drift-diffusion model.\cite{Peng43,Peng44,Peng45,Peng46,Peng47,Peng} 

In above sections, we have compared the analytical results [Eqs.~(\ref{Lsx}),
(\ref{Lox}), (\ref{Lly}), (\ref{Ltyminus}) and (\ref{Lty})]
 with the numerical ones.
 Now, we address the general condition that the analytical
  results can be applied. It is noted
  that our analytical results are derived in the strong scattering regime with
  $l_{\tau}\lesssim l_{\alpha},l_{\Omega}$ and then extended to the moderate
  scattering regime. The numerical calculations show that in the relatively
  strong and strong
  scattering regimes ($l_{\tau}\lesssim l_{\alpha},l_{\Omega}$), the analytical
  results agree with the numerical ones fairly well; in the
  moderate scattering regime, the condition to use the analytical results is
  $l_{\Omega}\lesssim l_{\tau}< l_{\alpha}$
  or $l_{\alpha}\lesssim l_{\tau}< l_{\Omega}$, which are close to the
  boundary between the moderate and
  relatively strong scattering regimes. However, even when the system is away
  from the
  boundary between the moderate and
  strong scattering regimes, the dependencies of the spin
    diffusion on the scattering strength,
    Zeeman field and SOC strength are qualitatively correct in the moderate
    scattering regimes.

\section{Conclusion and discussion}
\label{summary}
In conclusion, we have investigated the steady-state spin diffusion for the
  3D ultracold spin-orbit coupled
  $^{40}$K gas by the KSBE approach\cite{wu-review} first analytically and
then numerically. The spin diffusions along the $\hat{\bf x}$- and
$\hat{\bf y}$-directions for the transverse (${\bf P}||\hat{\bf z}$) and
longitudinal (${\bf P}||\hat{\bf x}$) configurations are studied. It is
first shown analytically that the behaviors of the steady-state spin diffusion 
 in the four configurations ($\hat{\bf x}$-T, $\hat{\bf x}$-L, $\hat{\bf y}$-T
 and $\hat{\bf y}$-L) 
are determined by three characteristic lengths: the mean free path $l_{\tau}$, the Zeeman oscillation
length $l_{\Omega}$, and the SOC oscillation
 length $l_{\alpha}$. We have derived the spin diffusion lengths for the
   spin diffusions in the four configurations in the strong scattering regime,
   which are then extended to the weak scattering one.
We further find that in different limits, the complex analytical reuslts can be
  reduced to different extremely simple forms, and correspodingly, the system can be
  divided into different regimes.
Specifically, it is revealed 
that by tuning the scattering strength, the system can be divided into {\it five}
regimes: I, weak
scattering regime ($l_{\tau}\gtrsim l_{\Omega}, 
l_{\alpha}$); II, Zeeman field-dominated moderate scattering regime ($l_{\Omega}\ll
l_{\tau}\ll l_{\alpha}$); III, SOC-dominated moderate scattering regime ($l_{\alpha}\ll
l_{\tau}\ll l_{\Omega}$); IV, relatively strong scattering regime ($l_{\tau}^c\ll l_{\tau}\ll l_{\Omega},
l_{\alpha}$); V, strong scattering regime ($l_{\tau}\ll l_{\Omega},
l_{\alpha}, l_{\tau}^c$). In different regimes, the corresponding behaviors of the spacial
 evolution of the spin polarization in the steady
state are very rich, showing different dependencies on the
scattering strength, Zeeman field and
SOC strength. These dependencies are summarized in
Table~{\ref{table_x}} (Table~{\ref{table_y}}) for the spin diffusion along the
$\hat{\bf x}$-direction ($\hat{\bf y}$-direction). Then the scattering strength, Zeeman field and
SOC strength dependencies of the spin diffusions are numerically calculated and
compared with the analytical
ones. It is shown that the analytical
  results agree with the numerical ones fairly well in the relatively strong/strong scattering
  regime and the region close to the
  boundary between the moderate and relatively
  strong scattering regimes. However, it is found that even when the system is away
  from the strong scattering regime, the analytical results are still qualitatively correct in the moderate
    scattering regimes.

The rich behaviors of the spin diffusions in different regimes
 are hard to be understood in the framework of the previous simple drift-diffusion
  model\cite{Peng43,Peng44,Peng45,Peng46,Peng47,Peng} or the direct inhomogeneous
  broadening [Eqs.~(\ref{diffusion_kx}) and (\ref{diffusion_ky})]
 picture\cite{2001,wu-review,diffusion_k,double,JinLuo_infinite} 
 in the literature.  
   In this work, we extend our previous inhomogeneous
  broadenings [Eqs.~(\ref{diffusion_kx}) and (\ref{diffusion_ky})] to the effective ones in
  Eqs.~(\ref{effective_Omega}) and (\ref{effective_alpha})
 [Eqs.~(\ref{effective_Omega_y}) and (\ref{effective_alpha_y})] for the spin
 diffusion along the $\hat{\bf x}$-direction
($\hat{\bf y}$-direction). In the limit situation, we suggest reasonable pictures referred to
as {\it modified} drift-diffusion model and {\it modified} inhomogeneous
broadening picture to facilitate the understanding of the simple analytical results in
Tables~\ref{table_x} and \ref{table_y}.
It is shown that the behaviors of the spin diffusions can be analyzed in the
  situation either with strong Zeeman and weak spin-orbit coupled fields
  (Regimes II and V) or weak Zeeman and strong spin-orbit coupled fields
  (Regimes III and IV).
When the spin
polarization is parallel (perpendicular) to the larger field between the Zeeman and
spin-orbit coupled fields, the spin polarization cannot (can) rotate around the effective
inhomogeneous broadening fields efficiently, and hence the
 {\it modified} drift-diffusion model ({\it modified} inhomogeneous
broadening picture) can be used. In the modified drift-diffusion model,
 in the strong scattering regime, $\tau_s({\bf k})$ remains the SRT in the
   conventional DP mechanism;\cite{relaxation_1,relaxation_2,relaxation_4,wu-review}
 whereas in the moderate scattering regime, $\tau_s({\bf k})$ is replaced by
 the helix spin-flip rates determined in the helix
  space.\cite{relaxation_3,Anomalous_DP_2} In the modified inhomogeneous
  broadening picture, the behavior of the spin diffusion is
determined by the {\it effective}
inhomogeneous broadenings together with the spin-conserving scattering. 
 Based on the modified drift-diffusion model
 and modified inhomogeneous broadening picture, apart from Regime
IV, all the features in
different regimes can be obtained. Below, we address several anomalous features of the spin
diffusion, which are in contrast to {\it both} the simple drift-diffusion
  model and the direct inhomogeneous
  broadening picture.

 In the scattering strength dependence, it is found that when $l_{\alpha}\ll l_{\Omega}$,
 the longitudinal spin diffusion along the $\hat{\bf y}$-direction is {\it
   robust} against the scattering even when the system is away from the strong scattering regime, which is in contrast
  to the simple drift-diffusion model. In that model, in the
  weak scattering regime, with $l_s\propto k\sqrt{\tau_k}/m$, the spin diffusion
  length is suppressed by the scattering. 
  In the Zeeman field
   dependence, when the system is in
Regime II, the {\it longitudinal} spin diffusion is enhanced by the Zeeman
field. This is in contrast to the prediction from the previous drift-diffusion model and
the direct inhomogeneous broadening picture. In the simple drift-diffusion model, in the strong (weak) scattering
regime, with $l_s\propto 1/(m|\alpha|)$ ($l_s\propto k\sqrt{\tau_k}/m$), it is obtained that the diffusion
length is irrelevant to the Zeeman field. In the direct inhomogeneous
broadening picture, the spin diffusion is suppressed due to the
  enhancement of the inhomogenous broadening by the Zeeman field.  
Finally, in the SOC strength dependence, we find that the spin diffusion length
can also be enhanced by the SOC in Regime III. This also goes beyond the
prediction from the simple drift-diffusion model and the direct inhomogeneous
broadening picture. In the simple drift-diffusion model, in the strong (weak)
  scattering regime, with $l_s\propto 1/(m|\alpha|)$ ($l_s\propto
  k\sqrt{\tau_k}/m$), the spin diffusion length is suppressed (uninfluenced) by
  the SOC. In the direct inhomogeneous
  broadenings picture, the spin diffusion is uninfluenced (suppressed) for the
spin diffusion along the $\hat{\bf x}$-direction ($\hat{\bf y}$-direction). All
these anomalous behaviors have been well understood from our modified
drift-diffusion model and/or modified inhomogeneous broadening picture.    

 We emphasize that for the longitudinal spin diffusion along
  the $\hat{\bf y}$-direction, the {\it robustness} against the
  scattering strength exists in a wide range including both the
 strong and {\it weak} scattering regimes. It is noted that in
    the strong scattering regime,  
    this feature has been
    predicted in the simple drift-diffusion model\cite{Peng43,Peng44,Peng45,Peng46,Peng47} and is also revealed
    in graphene by the KSBE approach.\cite{Peng} In this work, we further extend
    it into the weak scattering
  regime. Moreover, it is found that in a wide range of
the scattering,
 the corresponding diffusion length is only determined by the
    SOC strength, and hence also irrelevant to
    the atom density and temperature. One expects similar behavior in symmetric (110) quantum
 wells under a weak in-plane magnetic
field with similar SOC.\cite{Ohno110,Wu110,Dohrmann04,spin_noise110,Sherman110,Anomalous_DP_1,Anomalous_DP_2}
Moreover, the enhancement of the longitudinal spin diffusion by the Zeeman field has not
yet been reported in the literature, which is also expected to be observed in symmetric (110) quantum
 wells when the Zeeman energy larger than the spin-orbit coupled one.

Finally, we further compare the pictures of the spin diffusion provided in this
  work to understand the calculated results and those in the literature.
 In the previous
 works,
the spin diffusion in the {\it strong} scattering regime have been
extensively studied in the system with SOC, including
semiconductors,\cite{review1,Bloch,Awschalom,Zutic,fabian565,Dyakonov,wu-review,Korn}
 graphene\cite{Peng,nature,Peng32,Peng31,nano,Hanwei_review,recent} and recently
 monolayer MoS$_2$.\cite{Wang}
 Drift-diffusion
 model\cite{Peng43,Peng44,Peng45,Peng46,Peng47} and/or the inhomogeneous broadening
picture\cite{2001,wu-review,diffusion_k,double,JinLuo_infinite}
 were used to understand the behaviors of the spin diffusion.
In this work, the analytical results in the strong
 scattering regime are extended to the weak one and confirmed by the full numerical
 calculation.
For the strong scattering regime, we further divide it into
 the relatively strong and strong scattering regimes; whereas for the
 weak scattering regime, we divide it into the moderate and weak scattering regimes.   
We find all the anomalous
 behaviors revealed in this work appear in the moderate and relatively strong
 scattering regimes. Furthermore, our {\it
   modified} drift-diffusion model and/or {\it modified} inhomogeneous broadening
 picture are used to understand the behaviors of the spin diffusion in all these
 regimes. It is found that in the moderate and strong scattering regimes, these
 pictures work well. However, in the relatively strong scattering
 regime, which lies in the crossover of the moderate and strong scattering
 regimes, these pictures fail to explain the behaviors of the spin diffusion
 along the $\hat{\bf y}$-direction. This is because for
 the drift-diffusion model, in the relatively strong scattering regime, it is too rough to 
 consider the anisotropy between the diffusions along different
 directions.\cite{Peng} Nevertheless, when the scattering is strong enough, this
 anisotropy in the spin diffusion behavior is vanished. 
 Whereas for the inhomogeneous broadening picture,
 in this regime there exists strong competition
 between the effective inhomogeneous broadening and scattering, which makes the
 behavior of the spin diffusion complicated.\cite{Peng,JinLuo_infinite,double}

\begin{acknowledgments}

This work was supported
 by the National Natural Science Foundation of China under Grant
No. 11334014 and  61411136001, the National Basic Research Program
 of China under Grant No.
2012CB922002 and the Strategic Priority Research Program 
of the Chinese Academy of Sciences under Grant
No. XDB01000000.

\end{acknowledgments}

\begin{appendix}
\section{Analytical Analysis}
\label{AA}
We analytically derive the transverse and longitudinal spin diffusion lengths
for the spin diffusions along the $\hat{\bf x}$- and $\hat{\bf y}$-directions
 based on the KSBEs [Eq.~(\ref{KSBEs})]. 

Generally, the density matrix depends on both the zenith (between
${\bf k}$ and $\hat{x}$-axis) and azimuth (between ${\bf k}$ and $\hat{y}$-axis in the
$\hat{y}$-$\hat{z}$ plane) angles $\theta_{\bf k}$ and $\phi_{\bf k}$ in 3D. However, with the specific form of the
SOC [Eq.~(\ref{SOC})] and isotropic scattering terms [Eq.~(\ref{a_disorder})], we can define the quantity
\begin{equation}
\bar{\rho}_{{\bf k}}=\frac{1}{2\pi}\int_0^{2\pi}d\phi_{{\bf k}}\rho_{{\bf
    k}},
\label{order}
\end{equation}
which is averaged over the azimuth angle $\phi_{\bf k}$, to describe the kinetics
of the density matrix.\cite{relaxation_3}
 Accordingly, the KSBEs in the steady state 
become
\begin{eqnarray}
\nonumber
&&\frac{k_{\xi}}{m}\frac{\partial \bar{\rho}_{{\bf k}}({\bf r})}{\partial
  \xi}+i\Big[\Omega\sigma_x/2,\bar{\rho}_{{\bf
    k}}({\bf r})\Big]+i\Big[\alpha k_x\sigma_z/2,\bar{\rho}_{{\bf
    k}}({\bf r}) \Big]\\
&&\mbox{}+\sum\limits_{\bf k'}W_{\bf kk'}\big[\bar{\rho}_{\bf k}({\bf r})-\bar{\rho}_{\bf
  k'}({\bf r})\big]=0,
\label{KSBEs_averaged}
\end{eqnarray}
in which $\bar{\rho}_{{\bf k}}({\bf r})$ only depends on $\theta_{\bf k}$.

\subsection{Spin diffusion along the $\hat{\bf x}$-direction}
\label{AA1}
For the spin diffusion along the $\hat{\bf x}$-direction,
$\bar{\rho}_{\bf k}$ is expanded by the Legendre
function, which is written as
\begin{equation}
\bar{\rho}_{{\bf k}}=\sum_l\bar{\rho}_{k}^lC_l^0P_l(\cos{\theta_{{\bf k}}}),
\end{equation}
with $C_l^0=\sqrt{(2l+1)/(4\pi)}$. Accordingly, the dynamical equation for
 $\bar{\rho}_{k}^l$ is written as  
Eq.~(\ref{KSBE_x}). With the spin vector
defined by $\bar{{\bf S}}_{k}^l={\rm Tr}[\bar{\rho}_{k}^l{\bgreek \sigma}]$, the equations for
the spin vectors can be obtained. By further keeping the zeroth and first
  orders ($l=0,1$), the equation for the vector $\bar{{\bf
  S}}_k=(\bar{\bf S}_{k,x}^0,\bar{\bf S}_{k,y}^0,\bar{\bf S}_{k,z}^0,\bar{\bf
S}_{k,x}^1,\bar{\bf S}_{k,y}^1,\bar{\bf S}_{k,z}^1)^T$
 is written as 
\begin{equation}
\partial_x \bar{\bf S}_{k}+U_x\bar{\bf S}_{k}=0,
\end{equation}
with 
\begin{equation}
\hspace{-0.1cm}U_x=\left(
\begin{BMAT}[2.3pt,2pt]{cccccc}{cccccc}
0 & 1/l_{\alpha} & 0 & \sqrt{3}/l_{\tau} & 0 & 0\\
-1/l_{\alpha} & 0 & 0 & 0 & \sqrt{3}/l_{\tau} & 1/l_{\Omega}\\
0 & 0 & 0 & 0 & -1/l_{\Omega} & \sqrt{3}/l_{\tau}\\
0 & 0 & 0 & 0 & 1/l_{\alpha} & 0\\
0 & 0 & 1/l_{\Omega} & -1/l_{\alpha} & 0 & 0\\
0 & -1/l_{\Omega} & 0 & 0 & 0 & 0
\addpath{(2,3,.)rrrrrr}
\addpath{(3,3,.)dddddd}
\end{BMAT} \right).
\label{Ux}  
\end{equation}
From Eq.~(\ref{Ux}), the spin diffusion and oscillation lengths can be found from the
eigenvalues of $U_x$ denoted by $\lambda_x$, which satisfy 
\begin{equation}
\lambda_x^6+a\lambda_x^4
+b\lambda_x^2+c=0,
\label{eigen_x} 
\end{equation}
with $a=2/l_{\Omega}^2+2/l_{\alpha}^2$,
 $b=3/(l_{\tau}l_{\Omega})^2+1/l_{\Omega}^4+2/(l_{\Omega}l_{\alpha})^2+1/l_{\alpha}^4$
 and $c=-3/(l_{\tau}l_{\Omega}l_{\alpha})^2$.
In Eq.~(\ref{eigen_x}), the real and imaginary parts of $1/\lambda_x$ correspond
to the diffusion length and oscillation length, respectively.

 With $\Delta=(q/2)^2+(p/3)^3$ where $q=2a^3/27-ab/3+c$ and $p=b-a^2/3$, it is demonstrated that in the
  moderate and strong scattering regimes, $\Delta$ is always larger than
  zero. Therefore, there are one real root ($\Lambda_{\rm re}$) and two complex
  conjugate roots ($\Lambda_{{\rm im},\pm}$) for
  $\lambda_x^2$ in Eq.~(\ref{eigen_x}), which are written as
\begin{eqnarray}
&&\Lambda_{\rm re}=\sqrt[3]{-q/2-\sqrt{\Delta}}+\sqrt[3]{-q/2+\sqrt{\Delta}},\\
\nonumber
&&\Lambda_{{\rm im},\pm}=-(1/2)\Big(\sqrt[3]{-q/2-\sqrt{\Delta}}+\sqrt[3]{-q/2+\sqrt{\Delta}}\Big)\\
&&\mbox{}\pm (\sqrt{3}i/2)\Big(\sqrt[3]{-q/2-\sqrt{\Delta}}-\sqrt[3]{-q/2+\sqrt{\Delta}}\Big).
\end{eqnarray}
Accordingly, the spin diffusion length for the single exponential decay, the spin diffusion length for the
oscillation decay and the spin oscillation length are given by
\begin{eqnarray}
\hspace{-0.8cm}&&L_{s}^x=1/\sqrt{\Lambda_{\rm re}},
\label{Lsx}\\
\hspace{-0.8cm}&&L_{o}^x=\sqrt{2}{/}\sqrt{\Lambda_{\rm re}+\sqrt{\Lambda_{\rm
      re}^2+|\Lambda_{{\rm im},+}-\Lambda_{{\rm im},-}|^2/3}},
\label{Lox}\\
\hspace{-0.8cm}&&l_{o}^x=2\sqrt{3}L_{o}^x/|\Lambda_{{\rm im},+}-\Lambda_{{\rm
    im},-}|.
\label{llox}
\end{eqnarray}

\subsection{Spin diffusion along the $\hat{\bf y}$-direction}
\label{AA2}
For the spin diffusion along the $\hat{\bf y}$-direction, $\bar{\rho}_{\bf k}$
 is expanded by the Fourier 
function, which is denoted by
\begin{equation}
\bar{\rho}_{{\bf k}}=\sum_l\tilde{\rho}_{k}^l\exp(il\theta_{\bf k}).
\end{equation}
The corresponding dynamical equation for
 $\tilde{\rho}_{k}^l$ is written as  
Eq.~(\ref{KSBE_y}).
By keeping the zeroth and first orders ($l=0,1$), the equation for the vector ${\tilde{\bf
    S}}_k=\big(\tilde{\bf S}_{k,x}^0,\tilde{\bf S}_{k,x}^1-\tilde{\bf
  S}_{k,x}^{-1},
\tilde{\bf S}_{k,y}^0,\tilde{\bf S}_{k,z}^0,\tilde{\bf
  S}_{k,y}^1-\tilde{\bf S}_{k,y}^{-1},\tilde{\bf S}_{k,z}^1-\tilde{\bf S}_{k,z}^{-1}\big)^T$ is written as
\begin{eqnarray}
\partial_y \tilde{{\bf S}}_{k}+U_y\tilde{{\bf S}}_{k}=0,
\end{eqnarray}
where 
\begin{equation}
\hspace{-0.1cm}U_y=\left(
\begin{BMAT}[1pt,0.7pt]{cccccc}{cccccc}
0 & \frac{\displaystyle i}{\displaystyle l_{\tau}} & 0 & 0 & 0 & 0\\
\frac{\displaystyle
  -i}{\displaystyle l_{\tau}^2/(3l_{\Omega}^2)+1}\frac{\displaystyle l_{\tau}}{\displaystyle l_{\alpha}^2}&0 & 0 & 0 & 0 & 0\\
0 & 0 & 0 & 0 & \frac{\displaystyle i}{\displaystyle l_{\tau}} & \frac{\displaystyle i}{\displaystyle \sqrt{3}l_{\Omega}}\\
0 & 0 & 0 & 0 & -\frac{\displaystyle i}{\displaystyle \sqrt{3}l_{\Omega}} & \frac{\displaystyle i}{\displaystyle l_{\tau}}\\
0 & 0 & -\frac{\displaystyle il_{\tau}}{\displaystyle l_{\alpha}^2} &-\frac{\displaystyle 2i}{\displaystyle \sqrt{3}l_{\Omega}}  & 0 & 0\\
0 & 0 & \frac{\displaystyle 2i}{\displaystyle \sqrt{3}l_{\Omega}} & 0 & 0 & 0
\addpath{(2,4,.)rrrrrr}
\addpath{(2,3,.)dddddd}
\end{BMAT} \right).
\label{Uy}  
\end{equation}
In above equation, it is noted that the up-left $2\times 2$ and down-right
  $4\times 4$ blocks $U_y^{\rm u\mbox{-}l}$ and $U_y^{\rm d\mbox{-}r}$, which describe 
the spin diffusion for ${\bf S}_x$ and ${\bf S}_y$/${\bf S}_z$, are decoupled
to each other.

 Accordingly, from the
eigenvalues of $U_y^{\rm u\mbox{-}l}$, the spin diffusion length for ${\bf S}_x$ is
found to be 
\begin{equation}
L_{L}^y=l_{\alpha}\sqrt{l_{\tau}^2/(3l_{\Omega}^2)+1}.
\label{Lly}
\end{equation}
From $U_y^{\rm d\mbox{-}r}$, it
 is found that the four eigenvalues $\lambda_y$ satisfy 
\begin{equation}
\lambda_y^4+\big[4/(3l_{\Omega}^2)-1/l_{\alpha}^2\big]\lambda_y^2+4/(3l_{\Omega}^2l_{\tau}^2)+4/(9l_{\Omega}^4)=0.
\label{eigen_y}
\end{equation}
From Eq.~(\ref{eigen_y}), when
$1/l_{\alpha}^4-8/(3l_{\Omega}^2)(1/l_{\alpha}^2+2/l_{\tau}^2)\ge 0$, there are
four real roots, among which the two positive 
ones read 
\begin{equation}
|\lambda_y^{\pm}|=\sqrt{\frac{\displaystyle 1}{\displaystyle
    2l_{\alpha}^2}-\frac{\displaystyle 2}{\displaystyle 3l_{\Omega}^2}\pm
  \sqrt{\frac{\displaystyle 1}{\displaystyle 4l_{\alpha}^4}-\frac{\displaystyle
      2}{\displaystyle 3l_{\Omega}^2l_{\alpha}^2}-\frac{\displaystyle
      4}{\displaystyle 3l_{\Omega}^2l_{\tau}^2}}}.
\end{equation}  
Accordingly, the steady-state spin
polarization ${\bf S}_y$ or ${\bf S}_z$ is limited by the bi-exponential decay, with the
diffusion length being 
\begin{eqnarray}
&&L_{T}^{y,+}=1/|\lambda_y^+|,\\ 
&&L_{T}^{y,-}=1/|\lambda_y^-|,
\label{Ltyminus}
\end{eqnarray}
respectively. 
Otherwise, when
$1/l_{\alpha}^4-8/(3l_{\Omega}^2)(1/l_{\alpha}^2+2/l_{\tau}^2)<0$, the four
roots for $\lambda_y$ are complex. Specifically, the
real part of the roots for Eq.~(\ref{eigen_y}) is
identical,  which
is written as
\begin{equation}
\lambda_y^{\rm re}=\sqrt{1/(4l_{\alpha}^2)-1/(3l_{\Omega}^2)+\sqrt{1/(9l_{\Omega}^4)+1/(3l_{\Omega}^2l_{\tau}^2)}}.
\end{equation}
The complex part of the roots (absolute value) for Eq.~(\ref{eigen_y}) is
\begin{equation}
|\lambda_y^{\rm im}|=\sqrt{2/(3l_{\Omega}^2)(1/l_{\alpha}^2+2/l_{\tau}^2)-1/(4l_{\alpha}^4)}.
\end{equation}
Therefore, the steady-state spin polarization ${\bf S}_y$ or ${\bf S}_z$ is
determined by the oscillation
decay with the decay length and oscillation length being 
\begin{eqnarray}
&&L_{T}^y=1/\lambda_y^{\rm re},
\label{Lty}\\
&&l_{T}^y=2\lambda_y^{\rm re}/|\lambda_y^{\rm im}|,
\end{eqnarray}
respectively.

\end{appendix}

\end{document}